\documentclass[12pt]{article}
\usepackage{libertine}
\usepackage[margin=1.5in]{geometry}
\usepackage{newtxmath}
\usepackage[ruled,vlined, noend]{algorithm2e}
\usepackage{amsthm}
\usepackage{amsmath}
\usepackage{tikz}
\usetikzlibrary{decorations.pathreplacing, arrows}
\usetikzlibrary{shapes}
\allowdisplaybreaks[4]
\usepackage{graphicx}
\usepackage[font=sc]{caption}
\usepackage{subcaption}
\usepackage{booktabs}
\usepackage{threeparttable}
\usepackage{array, tabularx}
\usepackage{enumerate}
\usepackage{appendix}
\usepackage{indentfirst}
\usepackage{placeins}
\usepackage{dsfont}
\usepackage{adjustbox}
\usepackage{CJK}
\usepackage{bm}
\usepackage[T1]{fontenc}
\usepackage{listings}
\usepackage{epstopdf}
\usepackage{hhline}
\DeclareMathOperator*{\argmax}{argmax}

\usepackage{fancyhdr}
\usepackage{adjustbox}
\usepackage{multirow}
\usepackage[onehalfspacing]{setspace}
\usepackage{engord}
\usepackage{enumitem}
\usepackage{comment}

\usepackage{natbib}
\definecolor{winered}{rgb}{0.5,0,0}

\definecolor{lightgrey}{rgb}{0.95,0.95,0.95}
\definecolor{commentcolor}{RGB}{0,100,0}
\definecolor{frenchplum}{RGB}{190,20,83}

\RequirePackage{hyperref}
\hypersetup{
  pdfborder={0 0 0},
  colorlinks=true,
  linkcolor={blue},
  urlcolor={blue},
  filecolor={blue},
  citecolor={blue},
  linktoc=all,
}
\newcommand\Tstrut{\rule{0pt}{2.6ex}}         % = `top' strut
\newcommand\Bstrut{\rule[-0.9ex]{0pt}{0pt}}   % = `bottom' strut

\newtheorem{asu}{Assumption}
\newcounter{subassumption}[asu]
\renewcommand{\thesubassumption}{(\textit{\roman{subassumption}})}
\makeatletter
\renewcommand{\p@subassumption}{\theasu}% Counter prefix.
\makeatother
\newcommand{\subasu}{% Just like \item in a list, but for an asu
  \refstepcounter{subassumption}%
  \thesubassumption~\ignorespaces}

\newtheorem{definition}{Definition}
\newtheorem{lemma}{Lemma}
\newtheorem{proposition}{Proposition}
\newtheorem{corollary}{Corollary}

\newcommand*\pr{\text{Pr}}
\newcommand*\ex{\mathds{E}}

\newcommand*\ol{\overline}

\newcommand{\RNum}[1]{\uppercase\expandafter{\romannumeral #1\relax}}

\numberwithin{equation}{section}
\begin{document}

\title {\Large R\&D Heterogeneity and Countercyclical Productivity Dispersion\footnote{We are grateful to Christophe Chamley, Stephen Terry, Jianjun Miao and Robert G. King for guidance, encouragement and suggestions. For helpful comments and discussions, we thank Ryan Chahrour, Jean--Jacques Forneron, Loukas Karabarbounis, Stefania Garetto, Simon Gilchrist, Adam Guren, Tarek Hassan, Nyr Indictor, Danial Lashkari (discussant), Yuhei Miyauchi, Pierre Perron, Pascual Restrepo and participants in BU--BC Greenline Macro Meetings and numerous workshops.}}
\author{Shuowen Chen\thanks{Boston University. Email: swchen@bu.edu} \and Yang Ming\thanks{Central University of Finance and Economics. Email: mingyang@cufe.edu.cn}}
\date{\today}

\maketitle 
%\begin{center}
%\href{https://drive.google.com/file/d/1gX_g7oicMeeWBiTtMHNXAXdJhDx6ebr6/view?usp=sharing}{{\color{black}{Click Here for the Latest Version}}}
%\end{center}

\begin{abstract}
Why is the U.S. industry--level productivity dispersion countercyclical? Theoretically, we build a duopoly model in which heterogeneous R\&D costs determine firms' optimal behaviors and the equilibrium technology gap after a negative profit shock. Quantitatively, we calibrate a parameterized model, simulate firms' post--shock responses and predict that productivity dispersion is due to the low--cost firm increasing R\&D efforts and the high--cost firm doing the opposite. Empirically, we construct an index of negative profit shocks and provide two reduced--form tests for this mechanism.
\medskip

\end{abstract}
\par
%\textit{JEL Classification: O32 O33 O41 C73}
%\noindent \textit{Keywords:} Destructive innovation; Productivity dispersion; Technology ladder; R\&D
\newpage
\section{Introduction}\label{sec:intro}
Productivity dispersion, which measures the level--of--productivity difference between the most and the least productive groups of firms in an economy, has been found in the United States to be larger in recessions than in booms \citep{BLOOM2014, KEHRIG2015, BFJST2018}. In a seminal work, \cite{HK2009} argue that productivity dispersion reflects the degree of resource misallocation. This paper provides a new framework to explain the cause of this countercyclicality and illustrate that a greater dispersion in bad times can be rationalized by firms' optimal behaviors. The key mechanism is firms' heterogeneous R\&D responses to industry--level negative profit shocks.

%For U.S. manufacturing firms from 1970 to 2010, we construct an index of negative profit shocks as events for which the annual aggregate profit of an industry falls below the 5th percentile with respect to the overall profit. Using the local projection method \citep{JORDA2005}, we trace out the dynamic causal impacts of these shocks on industry--level productivity dispersion and R\&D intensity (RDI) dispersion. We find that both dispersions enlarge immediately at the onset of a negative profit shock and gradually dissipate after ten years. The findings are robust to three measures of dispersion: standard deviation, interquartile range (IQR), and difference between the 90th and 10th percentiles. 

%We start by an empirical investigation of the impact the negative profit shock to within-industry dispersion of productivity and R\&D intensity. Using the local projection method developed by \cite{JORDA2005}, and data from U.S. manufacturing firms, we find larger dispersion of both productivity and RDI in response to the shock at the industry level.

%The two new facts motivate our interpretation of countercyclical productivity dispersion as a result of heterogeneity in firms' R\&D activities. To understand how firms respond to negative profit shocks differently and the implications of this difference for the productivity dispersion, 
We develop a duopolistic model in which firms with heterogeneous R\&D costs participate in a dynamic game of stepwise innovations. In this model, a firm's profit increases in its technological advantage relative to its competitor. i.e., the technology gap. A firm improves its technology status via successful innovations, whose arrival rates are characterized by a Poisson process determined by the firm's R\&D efforts and an exogenous chance of catching--up if it is the laggard.  

%For analytical tractability, we impose a maximum technology gap and abstract from entry--exit dynamics. 
In the baseline model without a negative profit shock, we show that firms' optimal R\&D efforts and the stationary distribution of the technology gap are determined by firms' R\&D costs. More specifically, the firm with a lower marginal cost exerts more effort and is expected to be the leader in the stationary equilibrium. 
% , and the degree of such a recovery is determined by the technological progress of the leading firm

We then qualitatively analyze how firms respond to an exogenous negative profit shock of varying magnitudes. The model predicts that after a small shock, both firms increase their R\&D efforts to recover lost profits. For a large shock, however, both firms respond by reducing their R\&D efforts; they do this because they need many successful innovations to recover, but the increasing R\&D costs and risks of failures act as disincentives. These two extreme cases motivate our conjecture that for some shocks with a realistic magnitude, the leader responds with an increased R\&D effort while the laggard does the opposite. 

To quantitatively confirm this conjecture, we first construct an index of negative profit shocks using a newly merged panel dataset on U.S. manufacturing firms from 1970 to 2010. In this paper, an industry $j$ in year $t$ is hit by a negative profit shock if its detrended aggregate profit falls below the 5th percentile of the overall profit.\footnote{As robustness, we also consider using the 1st and 10th percentiles as thresholds.} Then, we calibrate and estimate a parameterized version of the model embedded in a general equilibrium framework {\`a} la \cite{AHHV2001}. We simulate firm responses by hitting the economy with a shock that decreases the industry's profit by 20\%. In response to this shock, the low--cost firm increases its R\&D effort on average while the high--cost firm does the opposite. Moreover, these responses translate into a hump--shaped impulse response curve of technology gap and generates countercyclical productivity dispersion. 

Turning to the empirical basis of this mechanism, we first use the within--industry empirical cumulative distribution functions (ECDF) of log R\&D intensities (RDI) as a proxy for firms' heterogeneity in R\&D responses: the lower its ECDF, the less active a firm is in R\&D. We show that, on average, firms with lower ECDFs in their previous--year RDI respond to a negative profit shock with lower current RDI, while firms with higher ECDFs in previous--year RDI do the opposite. 

We consider another specification that allows for nonlinear heterogeneity in R\&D responses. Firms in the same industry--year cell are sorted into ten decile groups by their log RDI, and the ECDF is replaced by dummy variables that indicate to which decile group a firm belongs. This approach shows that the least R\&D intensive firms respond to a negative profit shock by reducing RDI, while the most R\&D intensive firms do the opposite. Both empirical tests confirm the model prediction that heterogeneous R\&D activities in response to negative profit shocks leads to productivity dispersion at the industry level.
%%%%%%%%%%%%%%%%%%%%%%%%%%%%%%%%%%%%%%%%%%%%%%%%
\subsection*{Related Literature}
%First, we address the question of whether productivity dispersion necessarily implies an inefficient usage of factors. \cite{HK2009} argue that such a dispersion reflects the degree of resource misallocation and thus reduces economic welfare. \cite{KV2020} challenge this view by showing that an elimination of frictions like investment adjustment costs and external financing constraints can actually raise total output and productivity dispersion. We show that a greater dispersion in bad times can be rationalized by firms' optimal R\&D responses rather than institution--led misallocation. In other words, a greater productivity dispersion in bad times implies neither a decline nor an improvement in the efficiency of resources allocation. 
This paper contributes to three strands of literature, and this section provides a non--exhaustive review. First, we provide a new mechanism to the active literature on the cause of productivity dispersion \citep{Syverson2011, AghionAntoninBunel2021, AkcigitAtes2021, DeLockerSyverson2021}. In \cite{BM2011}, a negative aggregate profit shock induces firms to experiment with prices more drastically. This increases dispersion in consumer price and hence leads to a larger dispersion in measured revenue TFP. \cite{FGHW2019} and \cite{BLS2021} provide empirical support: a period of intensive innovations is accompanied by surge of entrants who engage in substantial experimentation and learning. These trials and errors yield different outcomes and thus induce dispersion in productivity. In \cite{KEHRIG2015}, a negative demand shock drives down factor prices. Hence more unproductive firms survive and productivity dispersion is enlarged. In \cite{Tian2015}, more firms take up riskier projects in recessions. Once successful, these projects bring out more output and enlarge productivity dispersion. 

Our mechanism does not feature frictions or entry--exit dynamics; instead, it is driven by firms' heterogeneous responses to shocks to profits: low-cost firms increase R\&D effort to recover the profit while high-cost firms are discouraged. Some recent papers propose explanations that also feature a reallocation of innovation activity from laggards to leaders. \cite{AghionBlundellGriffithHowittPrantl2009} provide empirical evidence that foreign firm entry induces heterogeneous responses of incumbents: for incumbents close to the tech frontier, the escape-entry effect incentivizes them to innovate more; for those far from the frontier, the discouragement effect dominates, and they shy away from innovation. In \cite{LMS2019}, leaders respond more aggressively than laggards to low interest rate because they want to escape neck--and--neck competition effects. Anticipating leaders' motive, laggards are discouraged and thus do not respond that much. In \cite{Olmstead-Rumsey2020}, productivity dispersion results from laggards failing to catch up with leaders in terms of the quality of innovations. \cite{AkcigitAtes2021} propose a mechanism in which a decrease of knowledge diffusion from leader to laggard enlarges the gap. 

% Laggards reduce R\&D effort due to the drop of gain from innovation. Leaders anticipate this and realize that their technology leaderships are safe, and thus discount future gains to innovation less and choose a slightly higher rate of innovations.
Second, we contribute to the theory of R\&D models with stepwise innovations. This literature dates back to \cite{TIROLE1988}, \cite{SAD1990}, \cite{GH1991}, \cite{AH1992} and \cite{BuddHarrisVickers1993}. Recent papers include \cite{AHHV2001}, \cite{ABBGH2005}, \cite{LS2016}, \cite{LMS2019} and \cite{AkcigitAtes2021}. Our model features one novelty: the incorporation of a negative profit shock. The analytical characterization only requires some general assumptions on the profit and R\&D cost functions, but we restrict the theoretical analysis to the maximum technology gap being one.\footnote{For tractability of the model, \cite{AkcigitAtes2021} also impose this assumption.} \cite{LMS2019} relax this assumption and allow firms to move apart indefinitely. Their analytical solutions, however, are predicated on the linearity of R\&D cost functions, and they resort to numerical analysis for general convex functions.\footnote{Convex R\&D cost implies positive R\&D by incumbents, and opens the door for strategic interaction among incumbents. On the other hand, linear innovation cost implies no R\&D by incumbents.} Our trade--off is that: we solve the model analytically by imposing a maximum gap but with abstract cost and profit functions. 

%The market structure assumed in the cited papers include a duopoly facing constant elasticity of substitution (CES) demand, a Cournot duopoly, and a Bertrand duopoly. 

Finally, we contribute to a burgeoning literature that uses tailed values of a distribution to define events. \cite{ACCS2016} regard a country as having a sovereign debt crisis if it experiences a quarterly change in spreads on sovereign debt over comparable risk--free debt that exceeds the top 5th percentile of the distribution of quarterly changes. \cite{AAC2014} define a patented innovation as being radical/disruptive if it is among the top 1\% patents based on citations received among all patents applied for in the same year. Many recent papers have adopted this approach, including \cite{CT2020}, \cite{CavenaileCelikTian2021} and \cite{CTW2021}. The shock we construct is exogenous to the firms, and we do not take a stand on its source. In addition, the choice of the threshold has an economic interpretation: the higher the threshold, the bigger magnitude of the shock. Our model predicts that firms are more responsive to a larger shock, and we provide empirical evidence from the data. 
%%%%%%%%%%%%%%%%%%%%%%%%%%%%%%%%%%%%%%%%%%%%%%%%
\subsection*{Structure of the Paper}
\indent The rest of the paper proceeds as follows: Section~\ref{sec:bslMdl} sets up a baseline model and establishes the optimal R\&D efforts and stationary distribution of the technology gap. Section~\ref{sec:extMdl} incorporates the negative profit shock to the baseline model and analyzes its effects on firms' R\&D efforts and the technology gap. Section~\ref{sec:empirics} introduces the datasets and discusses how we construct the negative profit shocks index. Section~\ref{sec:Quant} conducts quantitative analysis on a parameterized model and shows that a shock with a realistic magnitude induces productivity dispersion. Section~\ref{sec:emp} presents reduced--form evidence to the key mechanism. Section~\ref{sec:conc} concludes and discusses open questions. Details of data preprocessing, proofs and computations are available in Appendices \ref{asc:data}, \ref{Proof Appendix} and \ref{Computation Appendix} respectively. 
%%%%%%%%%%%%%%%%%%%%%%%%%%%%%%%%%%%%%%%%%%%%%%%%
%%%%%%%%%%%%%%%%%%%%%%%%%%%%%%%%%%%%%%%%%%%%%%%%
\section{Technology Ladders and Heterogeneous Climbers}\label{sec:bslMdl}
The main result of this section is that firms' optimal R\&D strategies endogenously generate a technology gap in the stationary equilibrium. Subsection~\ref{sec:setup} introduces the key ingredients. Subsection~\ref{basemodel:firm problem} describes firm's optimization problem. Subsection~\ref{subsec: MPE} characterizes the equilibrium and firms' optimal strategies. Subsection~\ref{subsec: Tech Gap Distribution} characterizes the limiting distribution of the technology gap. Subsection~\ref{subsec: Cost and Gap} discusses the distributions of technology gap under extreme scenarios of R\&D cost heterogeneity. 
%%%%%%%%%%%%%%%%%%%%%%%%%%%%%%%%%%%%%%%%%%%%%%%%
\subsection{Model Setup}\label{sec:setup}
%%%%%%%%%%%%%%%%%%%%%%
\noindent\textit{Technology Ladder and Maximum Gap.} -- Consider an industry with two firms $f\in\{A,B\}$. Time is continuous and there is a \textit{technology ladder} such that a firm that is further ahead on the ladder is more technologically advanced. Firm $f$'s location on the ladder at time $t$ is denoted by $n_{f}(t)$. The technology gap from firm $f$'s point of view at time $t$ is its distance from the rival on the ladder:  
$$\Delta n_{f}(t) = n_{f}(t) - n_{-f}(t),$$ 
where $-f$ denotes the rival of firm $f$. When $\Delta n_{f}(t)=0$, the two firms are neck--and--neck competitors. When $\Delta n_{f}(t)>0$, firm $f$ is the temporary leader while $-f$ is the laggard. From the laggard's point of view, the technology gap is negative. 

Can the technology distance between the two competing firms be infinitely large? It is reasonable to state that companies that have become very technologically distant are no longer considered to be competitors. For example, the horse-drawn vehicles and automobiles were once close competitors as means of transportation, but nowadays they are not considered as belonging to the same market. 

\begin{comment}
Figure (\ref{fig:modelintro}) illustrates firms' locations on the ladder at time $t$. 
\begin{figure}[ht] 
	\centering
	\caption{Model Description}
	\label{fig:modelintro}
	\begin{tikzpicture}[scale=0.8]
	[domain=-0.5:9]
	% Draw the two axes
	\draw[-] (0,0) -- (12,0) ;
	% Draw the four events on the x axis 
	\filldraw (0,0) circle(2pt) ;
	\filldraw (2,0) circle(2pt) ;
	\filldraw (2,0) circle(2pt) ;
	\filldraw (4,0) circle(2pt) node[below]{$n_{B}(t)$};
	\filldraw (6,0) circle(2pt) ;
	\filldraw (8,0) circle(2pt) node[below]{$n_{A}(t)$};
	\filldraw (10,0) circle(2pt) ; 
	\filldraw (12,0) circle(2pt) ;
	\draw [decorate,decoration={brace,mirror,amplitude=7pt},xshift=0pt,yshift=-1.3cm]
	(4,0.5) -- (8,0.5) node [below, midway, yshift = -0.4cm] {\footnotesize $\Delta (t)$};
	\draw [decorate,decoration={brace,amplitude=7pt},xshift=0pt,yshift=0pt]
	(0,0.5) -- (12,0.5) node [above, midway, yshift = 10pt] {\footnotesize $\text{Max Gap } \overline{m}$};
	\end{tikzpicture} 
\end{figure}
\end{comment}

We denote the maximum technology gap as $\ol{m}\in\mathds{N}_{+}$. To ensure the gap does not exceed this threshold at any $t$, we make the following assumption:

%%%%%%%%%%%%%%%
\begin{asu}[Automatic Catching Up] \label{asm:automaticgap}
When $\Delta n_{f}(t)=\overline{m}$ and the leader $f$ makes one innovation, the laggard $-f$ free rides the success: it moves one step forward automatically.
\end{asu} 
%%%%%%%%%%%%%%%

We provide two contexts to rationalize this assumption. First, some patents are expired and the leader no longer considers it as a threat that the laggard learns and imitates these old--fashioned technologies. Second, the laggard firm is left behind so much that it is replaced by a marginally better entrant who is just technologically one-step ahead of the laggard. This can be considered as a reduced--form way to represent firm entry--exit dynamics.\footnote{\cite{Olmstead-Rumsey2020} also imposes this assumption, but her model focuses on heterogeneous patent qualities rather than stepwise innovations. In her paper, the technology gap is used to parameterize firms' relative qualities of innovations.}

Assumption~\ref{asm:automaticgap} has an important implication: for all instants $t$, the set of technology gaps between the two firms, denoted by $\mathcal{M}$, has $(2\overline{m}+1)$ elements:
$$\mathcal{M}\coloneqq\{-\ol{m},-\ol{m}+1,\cdots,-1,0,1,\cdots,\ol{m}-1,\ol{m}\}.$$
%We show in Subsection~\ref{basemodel:firm problem} that $\mathcal{M}$ is the set of state space for the baseline model. 

Some previous papers also impose an upper bound on the gap to facilitate the theoretical analysis. For example, the maximum technology gap in \cite{AHHV2001} is implicitly determined by their restrictions on the stationary distribution. \cite{ABBGH2005} explicitly set the maximum gap to be one. In \cite{LS2016}, the maximum gap is implied under the assumption that firms stop doing R\&D at some point due to a strictly positive marginal cost of R\&D. One notable exception is \cite{LMS2019}, but as discussed in the literature review, they impose linear cost to obtain analytical solutions.
%%%%%%%%%%%%%%%%%%%%%%
\vskip 0.3cm
\noindent\textit{Profit Functions.} 
\begin{asu}[Firm Profit Function]\label{asm:pi}
    Denote $\boldsymbol{n}(t)=(n_{A}(t), n_{B}(t))$. The profit function $\pi_f\left(\boldsymbol{n}(t)\right)$ for  $f\in\{A,B\}$ satisfies the following properties: 
    \\
    \subasu \label{asm:pi1} 
    Time-invariancy: for any $\widetilde{\boldsymbol{n}}\in\mathds{N}_{+}^2$, and any $s,t\in[0,\infty)$, $\pi_f\left(\boldsymbol{n}(s)=\widetilde{\boldsymbol{n}}\right)=\pi_f\left(\boldsymbol{n}(t)=\widetilde{\boldsymbol{n}}\right)$.
    \subasu \label{asm:pi2}
    Monotonicity: $\pi_f\left(\boldsymbol{n}(t)\right)$ is strictly increasing in the technology gap $\Delta n_{f}(t)$.\\
    \subasu \label{asm:pi3}
    Symmetry: for any $n_1,n_2\in\mathds{N}_{+}$, $\pi_A\left(n_1,n_2\right)=\pi_B\left(n_2,n_1\right)$. 
\end{asu}

Assumption~\ref{asm:pi1} means that conditional on the technology gap, firm profit no longer depends on time index or individual productivity. Assumption~\ref{asm:pi2} means that a larger gap is associated with higher profits for the leader. As such, firm $f$ is motivated to innovate in order to escape competition with the  neck--and--neck rival \citep{AHHV2001}. By Assumption~\ref{asm:pi3}, conditional on the gap, firm profit does not depend on firm identity. 

We find it easier to prove theoretical results using the representation in Assumption~\ref{asm:pi}, but an equivalent expression of the profit function is $\pi_{f}(\Delta n_{f}(t))$. In Section~\ref{sec:extMdl}, we use the latter representation to incorporate the negative profit shock in the extended model. 

%Assumption~\ref{asm:pi} nests two leading cases in the literature on technology ladder models: the CSE production function in \cite{AHHV2001} and Bertrand duopoly in \cite{ABBGH2005}. However, Assumption~\ref{asm:pi2} doesn't hold for Cournot competition as in \cite{LS2016}, who assume that firm profit depends on the level of technology instead of the gap.
%%%%%%%%%%%%%%%%%%%%%%
\vskip 0.3cm
\noindent \textit{Uncertainty of R\&D}

Firms move up the ladder via innovations. To model the uncertainty of innovation, we assume that a success for firm $f$ at time $t$ follows a Poisson process with the  arrival rate 
\begin{equation} \label{eqn:RnD process}
    \lambda_f(t)=\lambda a_f(t)+h\cdot\mathds{1}\{n_f(t)<n_{-f}(t)\},
\end{equation}
where $a_f(t)\geq 0$ is the R\&D effort chosen by firm $f$ and the positive parameter $\lambda$ represents its effort efficiency. The greater the effort, the higher the probability that firm $f$ will improve its productivity relative to its competitor. The indicator function $\mathds{1}\{n_f(t)<n_{-f}(t)\}$ takes value one if $f$ is the laggard. The parameter $h$ denotes the exogenous rate of imitation for the laggard: with probability $h$, it can move one step ahead. This is an advantage the laggard has due to technological diffusion, which is a standard feature in the literature \citep{AHHV2001, LMS2019}. \cite{AkcigitAtes2021} consider this as a reduced--form representation of any mechanism that makes laggard learn from the leaders. 

The uncertain nature of R\&D implies that $n_{f}(t)$, firm $f$'s location on the ladder at time $t$, is random as well. More specifically, $n_{f}(t)$ follows a Poisson counting process $\left\{N_f(t): t>0\right\}$, in which random variables $N_f(t)$ and $N_{-f}(t)$ are independent for any $t>0$, except when their difference is $\overline{m}$. This is because by Assumption~\ref{asm:automaticgap}, the laggard moves up one step automatically if the leader makes an innovation. 

By Theorem 2.2 in \cite{GALLAGER1995}, the probability mass function (PMF) of firm $f$'s count of advancements along the technology ladder at a future instant $s$, conditional on its position at current instant $t$, takes the following form:

\begin{equation}\label{eqn:pmf}
    \mathds{P}{\left\{N_f(s)-N_f(t)=k \mid n_f(t)\right\}} = \dfrac{\left[\int_t^s \lambda_f(\tau)d\tau\right]^k \exp{\left[-\int_t^s \lambda_f(\tau)d\tau\right]}}{k!}, \quad k=0,1,\dots.
\end{equation}

This equation shows that the probability of firm $f$ moving up the ladder depends on the arrival rate of a successful innovation. Note that for a given $k$, this probability is not increasing in firm $f$'s effort because $\lambda_{f}$ also depends on the rival's decision. 
\begin{comment}
Figure (\ref{fig:statetrans}) illustrates the transition. 
\begin{figure}[ht] 
	\centering
	\caption{Transition of Technology on the Ladder}
	\label{fig:statetrans}
	\begin{tikzpicture}[scale=0.8, ->,>=stealth',auto,node distance=3cm,
  thick,main node/.style={circle,draw,font=\sffamily\Large\bfseries}]
	[domain=-0.5:9]
	% Draw the two axes
	\draw[-] (0,0) -- (12,0) ;
	% Draw the four events on the x axis 
	\filldraw (0,0) circle(2pt) ;
	\filldraw (2,0) circle(2pt) ;
	\filldraw (2,0) circle(2pt) ;
	\filldraw (4,0) circle(2pt) node[below] (1) {$n_{B}(t)$};
	\filldraw (6,0) circle(2pt) ;
	\filldraw (8,0) circle(2pt) node[below] (2) {$n_{A}(t)$};
	\filldraw (10,0) circle(2pt) node[below] (3) {$n_{B}(\textcolor{red}{s})$}; 
	\filldraw (12,0) circle(2pt) node[below] (4) {$n_{A}(\textcolor{red}{s})$};
    \path[every node/.style={font=\sffamily\small}]
    (1) edge[bend right=45] node [below] {$p_{B}^{st}(3)$} (3);
    \path[every node/.style={font=\sffamily\small}]
    (2) edge[bend left=90] node [above, midway] {$p_{A}^{st}(2)$} (4);
	\end{tikzpicture}
	\begin{minipage}{0.9\textwidth}
	{\footnotesize Note: In the left panel, $t^{-}$ denotes instant before the shock occurs. \par}
    \end{minipage}
\end{figure}
\end{comment}

%%%%%%%%%%%%%%%%%%%%%%
\vskip 0.3cm
\noindent\textit{Innovation Cost Functions} 

Doing R\&D incurs costs, and we impose the following assumption on the firm--specific cost function $\psi_f\left(a_f(t)\right)$.
\begin{asu}[Firm R\&D Cost Function] \label{asm:psi}
    For firm $f$ with effort $a_f(t)\geq 0$, the R\&D cost function $\psi_f(a_f(t))$ satisfies the following properties: \\
    \subasu \label{asm:psi1} $\psi_f(0)=0$; \\
    \subasu \label{asm:psi2} it is twice continuously differentiable; \\
    \subasu \label{asm:psi3}it is strictly increasing in $a_{f}(t)$; \\
    \subasu \label{asm:psi4}it is strictly convex in $a_{f}(t)$.
\end{asu}

Assumption~\ref{asm:psi1} means there is no fixed cost for R\&D, which is standard in Schumpeterian growth models. Assumption~\ref{asm:psi2} is a regularity condition to derive the model solution. Assumptions~\ref{asm:psi3} and \ref{asm:psi4} ensure that inputting more effort is more costly.   
%%%%%%%%%%%%%%%%%%%%%%%%%%%%%%%%%%%%%%%%
\subsection{Firm's Optimization Problem} \label{basemodel:firm problem}
A strategy profile for the firms in our model is 
$\left(\boldsymbol{a}(t)\right)_{t\geq 0}=\left(a_A(t),a_B(t)\right)_{t\geq 0},$ 
which specifies each firm's R\&D effort at time $t$. We define firm $f$'s \textit{performance measure}, namely its expected discounted sum of net profit flow, as follows:
\begin{equation}\label{eqn:pm}
    \mathcal{J}_f = \mathds{E}_{\boldsymbol{N}}\left\{\int_0^\infty e^{-\rho t}\left[\pi_f\left(\boldsymbol{N}(t)\right) - \psi_f\left(a_f(t)\right)\right]dt \mid \boldsymbol{n}(0),\left(\boldsymbol{a}(t)\right)_{t\geq 0}\right\},
\end{equation} 
where $\rho>0$ is the exogenous discount factor and the expectation is taken over the Poisson process $\left(\boldsymbol{N}(t)\right)_{t\geq 0}=\left(N_A(t),N_B(t)\right)_{t\geq 0}$, whose realization $\left(\boldsymbol{n}(t)\right)_{t\geq 0}=\left(n_A(t),n_B(t)\right)_{t\geq 0}$ determines the paths of the two firms on the technology ladder. The evolution of $\boldsymbol{N}(t)$ is governed by the equation~(\ref{eqn:pmf}) and the R\&D strategy profile $\left(\boldsymbol{a}(t)\right)_{t\geq 0}$, which also determines the arrival rates of innovations.

Given the strategy played by its competitor, firm $f$ chooses its own strategy to maximize~(\ref{eqn:pm}). The value function of $f$ is
\begin{equation}\label{eqn:valFcn}
    v_f\left(\boldsymbol{n}(0);\left(a_{-f}(t)\right)_{t\geq 0}\right) = \sup_{\left(a_f(t)\right)_{t\geq 0}} \mathds{E}_{\boldsymbol{N}}\left\{\int_0^\infty e^{-\rho t}\left[\pi_f\left(\boldsymbol{N}(t)\right) - \psi_f\left(a_f(t)\right)\right]dt\Big|\boldsymbol{n}(0),\left(\boldsymbol{a}(t)\right)_{t\geq 0}\right\}.
\end{equation}

This equation illustrates the two primary incentives behind a firm's strategy. Firstly, the firm bears the effort cost today but has a chance to improve its technology status by one step, which earns it higher profits in the future. Secondly, its expected gain today is also determined by how it expects the competitor to behave. The \textit{Nash equilibrium} (NE) of this game is a strategy profile in which neither firm has an incentive to deviate.

\begin{definition}[Nash Equilibrium] \label{def:spe}
    A Nash equilibrium is a pair of strategies $\left(a_A^*(t),a_B^*(t)\right)_{t\geq 0}$ such that given $\left(a_{-f}^*(t)\right)_{t\geq 0}$, the strategy $\left(a_f^*(t)\right)_{t\geq 0}$ is the solution to~(\ref{eqn:valFcn}).
\end{definition}

Equations~(\ref{eqn:pm}) and (\ref{eqn:valFcn}) are difficult to solve for because there are infinitely many state variables, namely instants of time. However, because the technology gap is a sufficient statistic for each firm's profit at time $t$ by Assumption~\ref{asm:pi}, the problem can be simplified so that the only state variable is the technology gap. The following lemma formally states the argument.

\begin{lemma}\label{lem:mkvStr}
    In a Nash equilibrium, suppose $s\geq 0$ and $t\geq 0$ satisfy $s\neq t$ and $n_{f}(s)-n_{-f}(s)=n_{f}(t)-n_{-f}(t)$. If $a_{-f}^*(s)=a_{-f}^*(t)$ and $a_f^*(t)$ is right-continuous, then $a_f^*(s)=a_f^*(t)$.
\end{lemma}

Lemma \ref{lem:mkvStr} states that if one firm plays an equilibrium strategy that is contingent on its technology gap, the other must do the same. More specifically, if $\left(a_{f}^*(t)\right)_{t\geq 0}$ is a \textit{Markov strategy}, in which action $a_{f}^*(t):=a_{f}^*\left(\Delta n_{f}(t)\right)$ depends solely on the state of technology distance $\Delta n_{f}(t)$, then the opponent's equilibrium strategy is Markovian as well. We focus on right-continuous strategies as they exclude cases in which there are discontinuity points of the first kind in the time paths of R\&D efforts, which is of no economic interest.

This reduction in the dimension of the strategy space leads to the \textit{Markov perfect equilibrium} (MPE) of our model, in which the policy and value functions are time invariant and only depend on the state variable.\footnote{Definition~\ref{def:mpe} is in line with the equilibrium concept in \cite{TIROLE1988} and \cite{MT2001}. We refer interested readers to \cite{MIAO2014} for more examples.} The rest of this paper focuses on MPE for analytical tractability.

\begin{definition}[Markov Perfect Equilibrium]\label{def:mpe}
    A Markov Perfect Equilibrium is a pair of state-contingent strategies $\left\{a_A^*(\Delta n_A),a_B^*(\Delta n_B)\right\}_{\Delta n_A=-\Delta n_B\in\mathcal{M}}$, where given $a_{-f}^*$, the strategy $a_f^*$ is the solution to firm $f$'s optimization problem:
    \begin{equation}\label{eqn:optMpe}
        \sup_{\left\{a_f(\Delta n_f)\right\}_{\Delta n_f\in\mathcal{M}}} \mathds{E}_{\Delta N_f}\left\{\int_0^\infty e^{-\rho t}\left[\pi_f\left(\Delta N_f(t)\right) - \psi_f\left(a_f\left(\Delta N_f(t)\right)\right)\right]dt \mid \Delta n_f(0),\left\{\boldsymbol{a}(\Delta n_f)\right\}\right\},
    \end{equation}
    where $\Delta N_{f}(t):= N_{f}(t) - N_{-f}(t).$
\end{definition}
%%%%%%%%%%%%%%%%%%%%%%%%%%%%%%%%%%%%%%%%%%%%%%%%
\subsection{Characterizing the MPE} \label{subsec: MPE}
Denote as $Z$ the time interval between the current instant and the instant in which an innovation arrives, and let $Z_f$ denote the case in which this innovation is conducted by firm $f$. By the memoryless property of equation (\ref{eqn:pmf}), firm $f$'s value function is
\begin{align}
    v_f\left(\Delta n_f\right) = &\mathds{E}_{Z} \bigg\{\int_0^Z e^{-\rho t}\left[\pi_f(\Delta n_f)-\psi_f\left(a_f^*(\Delta n_f)\right)\right]dt \nonumber \\
     &+ e^{-\rho Z}\mathds{1}\{Z=Z_f\}v_f\left(\Delta n_f + 1\right) + e^{-\rho Z}\mathds{1}\{Z=Z_{-f}\}v_f\left(\Delta n_f - 1\right)\bigg\}, \label{eqn:valFcnMpe}
\end{align}
where $\mathds{1}\{Z=Z_f\}$ takes one if the next innovation is achieved by firm $f$. The following proposition shows the existence of MPE and characterizes its form. 
\begin{proposition}[Existence and Characterization of MPE]\label{prp:exist}
    An MPE $a_f^*(\Delta n_f)$ for $f\in\{A,B\}$ and $\Delta n_f=-\Delta n_{-f}\in\mathcal{M}$ exists, and is the solution to the following optimization problem:
    \vskip 0.2cm
    
    \noindent $\forall \Delta n_f < \ol{m}$,
    \begin{align}
        v_f(\Delta n_f) & = \sup_{a_f\geq 0} \dfrac{1}{\lambda (a_f + a_{-f}^*) + h\cdot\mathds{1}\{\Delta n_f\neq 0\} + \rho}
        \Big\{\pi_f(\Delta n_f) - \psi(a_f) \nonumber \\
        &+ (\lambda a_f + h\cdot\mathds{1}\{\Delta n_f<0\})v_f(\Delta n_f + 1)  + (\lambda a_{-f}^* + h\cdot\mathds{1}\{\Delta n_f>0\})v_f(\Delta n_f - 1)\Big\}. \label{eqn:valFcnMPE} 
    \end{align}
    When $\Delta n_{f}=\overline{m},$
    \begin{align}
        &a_f^*(\ol{m}) = 0; \label{eqn:boundary}\\ 
        &v_f(\ol{m}) = \dfrac{1}{\lambda a_{-f}^* + h + \rho}
        \Big\{\pi_f(\ol{m}) + (\lambda a_{-f}^* + h)v_f(\ol{m} - 1)\Big\}. 
    \end{align}
\end{proposition}

In Proposition \ref{prp:exist} , $a_f^*$ and $v_f$ denote firm $f$'s policy function and value function in the MPE respectively. Equation~(\ref{eqn:valFcnMPE}) is the state-contingent value function transformed from equation~(\ref{eqn:optMpe}). Recall that we impose, in Assumption \ref{asm:automaticgap}, that the laggard free rides leader's innovation when the leader is at the maximum gap. It implies that the leader has no incentive to do any R\&D when the maximum gap is reached and thus the boundary condition (\ref{eqn:boundary}). 

The first order condition of equation  (\ref{eqn:valFcnMPE}) with respect to the R\&D effort implicitly determines the policy function $a_f^*$. Combined with firm $f$'s value function, we can characterize the model solution as the system of non-linear equations in the following corollary. Section \ref{sec:Quant} uses Corollary \ref{cor:mpe} to find the numerical solution to the model.

\begin{corollary}[Model Solution] \label{cor:mpe}
    The value functions $v_f(\Delta n_f)$ and policy functions $a_f^*(\Delta n_f)$ for $f\in\{A,B\}$ are solutions to the system of equations:
    \begin{align}  
        v_f(\Delta n_f) =& \dfrac{1}{\lambda \left[a_f^*(\Delta n_f) + a_{-f}^*(-\Delta n_f)\right] + h\cdot\mathds{1}\{\Delta n_f\neq 0\} + \rho}
        \Big\{\pi_f(\Delta n_f) - \psi_f\left(a_f^*(\Delta n_f)\right) & \nonumber \\
        &+ \left[\lambda a_f^*(\Delta n_f) + h\cdot\mathds{1}\{\Delta n_f<0\}\right]v_f(\Delta n_f + 1) & \nonumber \\
        &+ \left[\lambda a_{-f}^*(-\Delta n_f) + h\cdot\mathds{1}\{\Delta n_f>0\}\right]v_f(\Delta n_f - 1)\Big\}, \quad \text{if} \ \Delta n_f < \ol{m};&
        \label{eqn:valFcnBsl} \\
        \dfrac{d \psi_f\left(a_f^*(\Delta n_f)\right)}{d a_f} =& \lambda\left[v_f(\Delta n_f + 1) - v_f(\Delta n_f)\right], \qquad \qquad \qquad \qquad \quad \ \ \ \text{if} \ \Delta n_f < \ol{m};& \label{eqn:aFoc}
        \\
        a_f^*(\ol{m}) = &0; \label{eqn:aMax} &
        \\
        v_f(\ol{m}) = &\dfrac{1}{\lambda a_{-f}^* + h + \rho}
        \Big\{\pi_f(\ol{m}) + (\lambda a_{-f}^* + h)v_f(\ol{m} - 1)\Big\}. \label{eqn:vMax}&
    \end{align}
\end{corollary}

The denominator in firm $f$'s value function~(\ref{eqn:valFcnBsl}) contains both firms' optimal efforts. Intuitively speaking, the bigger the rival's effort, the less possible that the next innovation would be done by $f$. Equation~(\ref{eqn:aFoc}) conveys a clear message about how firm $f$'s R\&D effort is determined in equilibrium: the increment of firm value from an innovation, scaled by the marginal contribution of R\&D effort to the arrival rate of innovation, $\lambda$, should be equal to the marginal cost of R\&D. If the firm value increases more from innovation, or if the next innovation is expected to arrive sooner, then by strict convexity of the R\&D cost function, firm $f$ should exert a higher R\&D effort.

The system of equations in Corollary~\ref{cor:mpe} consists of $(8\ol{m}-4)$ equations and unknowns: $a_A^*$, $a_B^*$, $v_A$ and $v_B$ for $(2\ol{m}-1)$ states. By Proposition~\ref{prp:exist}, the MPE exists, and so does the solution to this system. Establishing the uniqueness is theoretically involved, but in the quantitative section, we do not have the issue of multiple equilibria. 
%%%%%%%%%%%%%%%%%%%%%%%%%%%%%%%%%%%%%%%%%%%%%%%%
\subsection{The Stationary Distribution of the Technology Gap} \label{subsec: Tech Gap Distribution}
The technology gap follows an endogenous Markov process whose transition rate is governed by firms' effort decisions. Recall that $\mathcal{M}$, the set of technology gaps, has $(2\ol{m}+1)$ components. We define $q_{ij}$, the \textit{transition rate} of the gap from the state $\mathcal{M}_i$ to an adjacent state $\mathcal{M}_j$, as follows:

\begin{equation}\label{eqn:eleQ}
    q_{ij} = 
    \begin{cases}
        \lambda a_A^*\left(\mathcal{M}_i\right) + h\cdot\mathds{1}\{\mathcal{M}_i<0\}, & \quad \text{if} \quad j = i + 1; \\
        \lambda a_B^*\left(-\mathcal{M}_i\right) + h\cdot\mathds{1}\{\mathcal{M}_i>0\}, & \quad \text{if} \quad j = i - 1.
    \end{cases}
\end{equation}
%Following the continuous--time Markov chain models literature \citep{NORRIS1998}, we do not consider the rival not succeeding as it is a zero measure event in any instant.
In our model, a firm can only jump one unit per innovation, which implies that the laggard cannot leapfrog the leader.\footnote{This is the key assumption in \cite{LMS2019} and they provide an extensive discussion on the empirical support and restrictions.}  Intuitively speaking, when $|i-j|=1$, the arrival time of an innovation by firm $f$, $Z_f$, follows an exponential distribution with rate $q_{ij}$. This implies that the expected time for the gap to jump to an adjacent state is $1/q_{ij}$. As a property of the sum of independent Poisson processes, the rate at which the technology gap leaves its current state $\mathcal{M}_i$ is hence $q_i\coloneqq q_{i,i+1}+q_{i,i-1}.$

\begin{figure}[!htbp]
    \centering
    \caption{Transition rates across states of technology gap}\label{fig:markov}
    \includegraphics[width=0.75\textwidth]{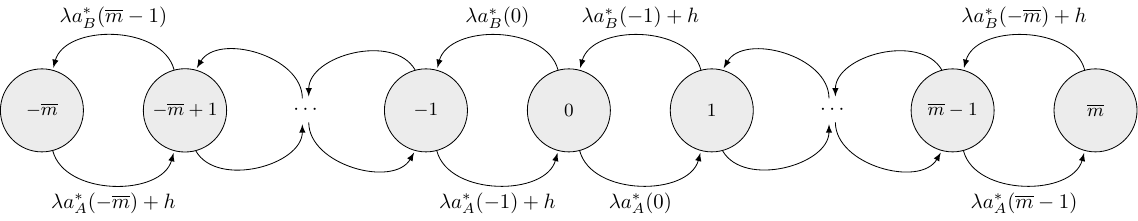} 
    %\begin{minipage}{0.75\textwidth}
	%{\footnotesize Note: \textcolor{red}{Add some description of the graph. Take three states ($-1, 0, 1$) as example and explain what the notations mean.}. \par}
    %\end{minipage}
\end{figure}

Now we define the transition matrix $Q$, which describes the probability of each state jumping to its adjacent states:
\[
Q_{ij}=
\begin{cases}
    q_{ij} & |i-j|=1, \\
    -q_{i} & |i-j|=0, \\
    0 & |i-j|>1.
\end{cases}
\]
%The last case is because it is impossible for the technology gap to jumps with size greater than one. 
%The Q--matrix allows to infer the distribution of the time between two jumps onto certain states. 
Figure (\ref{fig:markov}) is a graphical illustrates of the $Q$-matrix. As an example, state 0 means the two firms are neck--and--neck. By equation~(\ref{eqn:RnD process}), firm $A$ has an innovation with probability $\lambda a^{*}_{A}(0)$, which is thus the probability that the gap jumps to state 1. In a similar vein, the gap jumps to $-1$ if firm $B$ makes an innovation, which occurs with probability $\lambda a^{*}_{B}(0)$. In state 1, the gap jumps to state 0 if firm $B$ makes an innovation, which occurs with probability $\lambda a^{*}(-1)+h$ as $B$ is the laggard and the technology gap from its point of view is -1. 

Characterizing the limiting distribution of the technology gap requires knowing the transition rate between any two states. This information can be recovered using the Q--matrix:
\begin{definition}\label{def:matP}
    The jump matrix $P$ summarizes the transition rate between any two state and takes the form
    \begin{equation}\label{eqn:matP}
        P = \left(p_{ij}: i,j\in \{1,\cdots,2\ol{m}+1\}\right),
    \end{equation}
    where
    \begin{align}\label{eqn:eleP}
        p_{ij} =& 
        \begin{cases}
            \frac{-q_{ij}}{q_{ii}}, & \quad \text{if} \quad i\neq j, q_{ii}\neq 0; \\
            0, & \quad \text{if} \quad i\neq j, q_{ii}=0.
        \end{cases}
        \\
        p_{ii} =& 
        \begin{cases}
            0, & \quad \text{if} \quad q_{ii}\neq 0; \\
            1, & \quad \text{otherwise}.
        \end{cases}
    \end{align}
\end{definition}
\noindent We now define the \textit{stationary distribution} of the technology gap $\mu(m)$ for $m\in\mathcal{M}$.
\begin{definition}[Stationary Distribution of Technology Gap]\label{def:statDist}
    A length $(2\ol{m}+1)$ vector $\mu$ is the stationary distribution over state space $\mathcal{M}$ if and only if: \\
    {\normalfont (i)} $\mu P = \mu$ and {\normalfont (ii)} $\sum_{m\in\mathcal{M}} \mu(m) = 1$.
\end{definition}
\noindent By Theorem 5.11 in \cite{CINLAR1975}, we can consider $\mu$ as the \textit{limiting distribution} because

\begin{equation}\label{eqn:limDist}
    \mu(m) = \lim_{t\rightarrow \infty}\mathds{P}{\left(\Delta n_{f}(t)=m\right)}, \quad \forall m\in\mathcal{M}.
\end{equation}

The equation says that when $t$ is sufficiently large, the probability that the technology gap takes a certain value $m$ does not depend on the initial gap. Therefore, the limiting distribution $\mu$ characterizes the long-term behavior of the technology gap. 

The stationary distribution exists and is unique if the associated Markov chain of matrix $P$ is irreducible and recurrent. Figure~(\ref{fig:markov}) provides a clear illustration that these conditions are satisfied. Moreover, the stationary distribution can be explicitly expressed as follows:

\begin{proposition}[Analytical Form of the Stationary Distribution]\label{prp:statDist}

In the MPE, the stationary distribution $\mu$ of technology gap $\left(\Delta n_{f}(t)\right)_{t\geq 0}$ exists and is unique. More specifically, $$\mu(m)=-\xi_iq_{ii},$$ 
where $m=\mathcal{M}_i$ and $\xi_i$ is the $i$-th component of the solution to $\xi Q=0$.
\end{proposition}

Proposition~\ref{prp:statDist} is useful for the quantitative analysis: once we find a solution to equilibrium R\&D strategies $\left\{a_A^*(\Delta n_A),a_B^*(\Delta n_B)\right\}_{\Delta n_f\in\mathcal{M}}$, the stationary distribution can be calculated. In Section~\ref{sec:Quant}, we use $\mu$ to calculate moments of variables in the MPE. 
%%%%%%%%%%%%%%%%%%%%%%%%%%%%%%%%%%%%%%%%%%%%%%%%
\subsection{R\&D Costs and Expected Technology Gap} \label{subsec: Cost and Gap}
Armed with the limiting distribution of technology gap, we discuss the expected technology gap under two scenarios: (1) firms having identical R\&D cost functions, (2) one firm's marginal cost being pointwise lower than that of the other. The following proposition states that when two firms have the same cost function, the expected technology gap under the limiting distribution is zero.
\begin{proposition}\label{prp:gapHomo}
    If $\psi_A(a)=\psi_B(a)$ for all $a\geq 0$, then $\displaystyle \lim_{t\rightarrow \infty}\mathds{E}\left[\Delta n_{f}(t)\right]=0$.
\end{proposition}

Heuristically, this is not surprising as the two firms are identical in every aspect, which implies their optimal strategies coincide. Hence there is no reason for one firm to be the leader in the long term.
%In other words, firm $A$ is more efficient in R\&D than firm $B$. 

The more interesting case is in which firms have heterogeneous R\&D costs. Without loss of generality, assume that firm $A$'s marginal R\&D cost is strictly lower than that of firm $B$ at all R\&D effort levels. The following lemma shows that firm $A$'s value function exceeds its rival's over all possible states.

\begin{lemma}\label{lem:valFcn}
    If $\dfrac{\partial\psi_A(a)}{\partial a_A}<\dfrac{\partial\psi_B(a)}{\partial a_B}$ for all $a>0$, then for any $m\in\mathcal{M}$, $v_A(m)>v_B(m)$.
\end{lemma}

\noindent The following lemma shows that firm $A$'s R\&D effort dominates firm $B$'s at all states. 
\begin{lemma}\label{lem:rdi}
    Assume the following three conditions hold:
    \vskip 0.2cm
    \indent {\normalfont{(i)}} $\ol{m}=1$,
    \vskip 0.3cm
    \indent {\normalfont{(ii)}} $\dfrac{\partial\psi_A(a)}{\partial a_A}<\dfrac{\partial\psi_B(a)}{\partial a_B}$ for all $a>0$,
    \vskip 0.3cm
    
    \indent {\normalfont{(iii)}} The discount factor $\rho$ and the arrival rate multiplier $\lambda$ are sufficiently small.
    \vskip 0.3cm
    
    Then $a_A^*(m)>a_B^*(m)$ for $m=-1,0$.
\end{lemma}

Due to complexity of the non-linear system in Corollary \ref{cor:mpe}, we are yet able to generalize the proof to $\overline{m}>1$.%\footnote{\textcolor{blue}{
% YM008-JEDC
%The complexity is also responsible for condition (iii) in Lemma \ref{lem:rdi}, which is technically sufficient but not necessary.}} 
In the meantime, the qualitative restriction of parameters $\rho$ and $\lambda$ in Lemma \ref{lem:rdi} reflects the fact that we do not impose specific functional forms on profits and R\&D costs. In Section \ref{sec:Quant}, however, we show that the quantitative version of the model provides numerical values of $\rho$ and $\lambda$, and extends the statement to $\overline{m}>1$.

Naturally, the difference in optimal R\&D efforts leads to asymmetric probability masses on state space $\mathcal{M}$ in the long run. The following proposition states that in the limiting distribution, the low-cost firm has a greater chance of being the leader.
\begin{proposition}\label{prp:asymProb}
    Under the conditions in Lemma \ref{lem:rdi}, in the limiting distribution, the probability mass function of technology gap satisfies $\mu(1)>\mu(-1)$.
\end{proposition}

\noindent The proposition implies the following corollary that the expected technology gap is nonzero. 

\begin{corollary}[Expected Non--Zero Technology Gap]\label{cor:gapHeter}
    Under the conditions in Lemma \ref{lem:rdi}, $$\displaystyle \lim_{t\rightarrow \infty}\mathds{E}\left[\Delta n_{A}(t)\right]>0.$$
\end{corollary}

To summarize, this section presents the baseline model without a negative profit shock. The key prediction is that, when firms have heterogeneous R\&D costs, the expected technology gap is non-zero: the lower--cost firm conducts more R\&D in equilibrium. 
%%%%%%%%%%%%%%%%%%%%%%%%%%%%%%%%%%%%%%%%%%%%%%%%
%%%%%%%%%%%%%%%%%%%%%%%%%%%%%%%%%%%%%%%%%%%%%%%%
\section{Extended Model with a Negative Profit Shock}\label{sec:extMdl}
In this section, we incorporate a negative profit shock into the model and examine its impact on optimal R\&D efforts and technology gaps in equilibrium. 
% The baseline model in Section~\ref{sec:bslMdl} features only one state variable, i.e., the technology gap, whose expected value in the limiting distribution depends on firms' R\&D costs.
%%%%%%%%%%%%%%%%%%%%%%%%%%%%%%%%%%%%%%%%%%%%%%%%
\subsection{Modelling the Shock}
In this paper, a negative profit shock is unanticipated and affects both firms equally. Depending on the magnitude of the shock, firms have two different incentives. On one hand, if the shock is mild, then firms have an incentive to conduct R\&D to recover the profit. On the other hand, a devastating shock requires many successful innovations. Firms are not sure if their efforts will lead to successes, but the costs are for sure. Combined with already--low post--shock profits, firms are discouraged from conducting R\&D. 

Because we assume that a negative profit shock affects both firms equally, the technology gap is not affected. This implies that the profit function in the baseline model is not affected by the shock either. To model firms' reactions after the shock, we hence modify the profit function. More specifically, we introduce the concept of \textit{distance from the technology frontier} to operationalize firms' incentive to innovate. We first formalize the notion of a negative profit shock in the model as follows:
\begin{definition}[Negative Profit Shock]\label{def:inovShk}
    Denote as $d_f(t)$ firm $f$'s distances to the technology frontier at time $t$. A negative profit shock with magnitude $D\in\{1,\cdots,\ol{D}\}$ occurs at time $t$ if and only if:
    \vskip 0.2cm
    \indent {\normalfont{(i)}} There exists some $\tau>0$ such that for any $s\in(t-\tau,t)$, $\min\{d_A(s),d_B(s)\}=0$; 
    \vskip 0.2cm
    \indent {\normalfont{(ii)}} For firm $f\in\{A, B\}$, $d_f(t) - \displaystyle\lim_{s\rightarrow t^{-}}d_f(s)=D$.
\end{definition} 

\begin{figure}[ht] 
	\centering
	\caption{Graphical Illustration of a Negative Profit Shock} \label{fig:withNPS}
	% The instant before NPS hits
	\begin{tikzpicture}[scale=0.5]
	[domain=-0.5:9]
	% Draw the two axes
	\draw[-] (0,0) -- (12,0) ;
	% Draw the four events on the x axis 
	\filldraw (0,0) circle(2pt) ;
	\filldraw (2,0) circle(2pt) ;
	\filldraw (2,0) circle(2pt) ;
	\filldraw (4,0) circle(2pt) ;
	\filldraw (6,0) circle(2pt) ;
	\filldraw (8,0) circle(2pt) node[below]{$n_{B}(t^{-})$};
	\filldraw (10,0) circle(2pt) ; 
	\filldraw (12,0) circle(2pt) node[below]{$n_{A}(t^{-})$};
	\draw [decorate,decoration={brace,amplitude=7pt},xshift=0pt,yshift=0pt]
	(8,0.5) -- (12,0.5) node [above, midway, yshift = 10pt] {\footnotesize $\Delta n_{A}(t^{-})$};
	\end{tikzpicture}
	\qquad % <----------------- SPACE BETWEEN PICTURES
	% The instant at which NPS hits
	\begin{tikzpicture}[scale=0.5]
	[domain=-0.5:9]
	% Draw the two axes
	\draw[-] (0,0) -- (12,0) ;
	% Draw the four events on the x axis 
	\filldraw (0,0) circle(2pt) ;
	\filldraw (2,0) circle(2pt) ;
	\filldraw (2,0) circle(2pt) ;
	\filldraw (4,0) circle(2pt) node[below]{$n_{B}(t)$};
	\filldraw (6,0) circle(2pt) ;
	\filldraw (8,0) circle(2pt) node[below]{$n_{A}(t)$};
	\filldraw (10,0) circle(2pt) ; 
	\filldraw (12,0) circle(2pt) node[below]{$\text{Frontier}$};
	\draw [decorate,decoration={brace,amplitude=7pt},xshift=0pt,yshift=0pt]
	(8,0.5) -- (12,0.5) node [above, midway, yshift = 10pt] {\footnotesize $d_{A}(t)=D$};
	\draw [decorate,decoration={brace,amplitude=7pt},xshift=0pt,yshift=0pt]
	(4,0.5) -- (8,0.5) node [above, midway, yshift = 10pt] {\footnotesize $\Delta n_{A}(t)$};
	\draw [decorate,decoration={brace, amplitude=7pt},xshift=0pt,yshift=2cm]
	(4,0.5) -- (12,0.5) node [below, midway, yshift = 25pt] {\footnotesize $d_{B}(t)$};
%	\draw [red, decorate,decoration={brace,mirror, amplitude=7pt},xshift=0pt,yshift=-2cm]
%	(4,0.5) -- (12,0.5) node [below, midway, yshift = -10pt] {\footnotesize $d_{B}(t)$};
	\end{tikzpicture}
	\begin{minipage}{0.9\textwidth}
	{\footnotesize Note: In the left panel, $t^{-}$ denotes instant before the shock occurs. The technology gap between the two firms are not affected by the shock, i.e., $\Delta n_{A}(t^{-})=\Delta n_{A}(t)$. For notation, $\Delta n_{A}(t) = -\Delta n_{B}(t)$.\par}
    \end{minipage}
\end{figure}
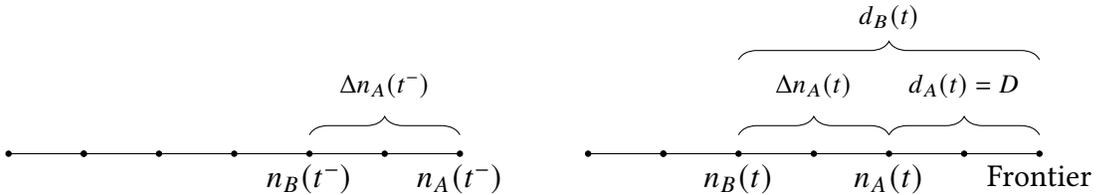

Figure (\ref{fig:withNPS}) illustrates how a negative profit shock affects the firms in our model. At instant $t^{-}$ before the shock, the frontier is just the the leader's (firm $A$) technology level, i.e., $\min\{d_{A}(t^{-}), d_{B}(t^{-})\}=0$. When a shock with magnitude $D$ occurs at time $t$, it pushes both firms off the technology frontier immediately by $D$ steps, and $\min\{d_{A}(t), d_{B}(t)\}=D$. 

The profit function now takes the form in Assumption~\ref{asm:piExt}. Both firms' profits are lowered depending on how far away the leader is from the frontier. 

\begin{asu}[Profit Function in the Model with NPS]\label{asm:piExt}
   Given distances to the technology frontier $d_f(t)$ and $d_{-f}(t)$, the profit of firm $f$ is
   \begin{equation}\label{eqn:piExt}
    \widetilde{\pi}_f(d_f(t),d_{-f}(t)) = (1-\delta D)\cdot\pi(d_{-f}(t)-d_f(t)),
   \end{equation}
   where $\delta\in(0,1)$ controls the severity of one degree of the shock, and the function $\pi(\cdot)$ satisfies Assumption \ref{asm:pi} in Section~\ref{sec:bslMdl}.
\end{asu}

The parameter $\delta$ controls the severity of one degree of the shock. It also determines the maximal magnitude of shock $\ol{D}=\max\{n\in\mathds{N}:\delta n \leq 1\}$. 

% SC: incorporated YM004
We make four remarks on our modelling approach. First, although a negative profit shock pushes firms off technology frontier, this does not mean that the firms have lost their knowledge. It only reflects that due to the negative profit shock, firms need higher levels of productivity to regain the pre-shock level of profit. Second, the technology gap is now the difference between leader's and laggard's distance to the frontier. This implies that the extended model now has two state variables, i.e., firms' respective distances to the frontier. The next subsection discusses cases of state transitions. 
% revision mark: YM004-JEDC
If the shock were anticipated, then the remaining time to the shock getting realized became another state variable and would make the model intractable. Third, it is empirically difficult to distinguish between the degree of shock, $D$, from the effect of one unit of shock $\delta$. In our empirical tests, we define the shock as a large drop in profit at the industry level. We provide more discussion in Section~\ref{sec:Quant}. 

%%%%%%%%%%%%%%%%%%%%%%%%%%%%%%%%%%%
\subsection{State Transitions}
Since the state space is now two-dimensional, the transition of state $(d_A(t),d_B(t))$ is more complex than in the baseline model. More specifically, there are four possible transitions after the arrival of an innovation. 

\textbf{Case 1:} $d_f(t)=d_{-f}(t)=0$: there is no shock and the two firms are neck--and--neck on the technology frontier. If the next innovation is made by firm $f$ at time $s$, then $d_f(s)=0$ and $d_{-f}(s)=1$, meaning that firm $f$ becomes the leader and stays on the frontier.

\textbf{Case 2:} $d_f(t)=0$ and $0<d_{-f}(t)<\ol{m}$: there is no shock and firm $f$ is the leader. If the next innovation is done by the leader at time $s>t$, then $d_f(s)=0$ and $d_{-f}(s)=d_{-f}(t)+1$. If it is the laggard who innovates, $d_f(s)=0$ and $d_{-f}(s)=d_{-f}(t)-1$.

\textbf{Case 3:} $d_f(t)=0$ and $d_{-f}(t)=\ol{m}>0$. The only difference from case 2 is that the leader $f$ is at the maximum technology gap. By Assumption~\ref{asm:automaticgap}, it has no incentive to do R\&D. Therefore, the next innovation must be from the laggard.

\textbf{Case 4:} $d_f(t)>0$ and $d_{-f}(t)>0$: there has been a shock at time $t$ or later. If the next innovation is from firm $f$ at time $s>t$, then $d_f(s)=d_f(t)-1$. For the other firm $-f$, if $d_{-f}(t)=d_f(t)+\ol{m}$, then $d_{-f}(s)=d_{-f}(t)-1$ due to Assumption~\ref{asm:automaticgap}; otherwise it remains where it was: $d_{-f}(s)=d_{-f}(t)$.

To formalize the discussion above, let $\left(d_f(t),d_{-f}(t)\right)$ denote the state of firm $f$ and its rival at time $t$. Suppose the next innovation arrives at time $s>t$, and is conducted by firm $f^\prime\in\{f,-f\}$.\footnote{We do not discuss the case in which both firms innovate simultaneously, as such an event occurs with zero probability.} Then the new state $\left(d_f(s),d_{-f}(s)\right)$ is governed by the transition rule $T$ as a function of $\left(d_f(t),d_{-f}(t)\right)$ and $f^\prime$.
\begin{equation}\label{eqn:transRule}
    T\left(d_f,d_{-f} \mid f^\prime\right) =
    \begin{cases}
        \left(d_f,d_{-f}+1\right), & \text{if} \ f^\prime=f, \quad d_f=0, \quad 0\leq d_{-f} <\ol{m}; \\
        \left(d_f-1,d_{-f}\right), & \text{if} \ f^\prime=f, \quad d_f>0, \quad d_f-d_{-f}>-\ol{m}; \\
        \left(d_f-1,d_{-f}-1\right), & \text{if} \ f^\prime=f, \quad d_f>0, \quad d_f-d_{-f}=-\ol{m}; \\
        \left(d_f+1,d_{-f}\right), & \text{if} \ f^\prime=-f, \quad d_{-f}=0, \quad 0\leq d_f <\ol{m}; \\
        \left(d_f,d_{-f}-1\right), & \text{if} \ f^\prime=-f, \quad d_{-f}>0, \quad d_{-f}-d_{f}>-\ol{m}; \\
        \left(d_f-1,d_{-f}-1\right), & \text{if} \ f^\prime=-f, \quad d_{-f}>0, \quad d_{-f}-d_{f}=-\ol{m}.
    \end{cases}
\end{equation}
As such, the rule $T$ returns the updated state upon the arrival of innovation depending on the current state and who the innovator is.

%As in the baseline model, we allow for knowledge diffusion $h$, with which the gap between the two firms shrinks by one step at a fixed rate $h\geq 0$ regardless of firms' R\&D efforts. 
%%%%%%%%%%%%%%%%%%%%%%%%%%%%%%%%%%%%%%%%%%%%%%%%
\subsection{Analysis of the Equilibrium}
Given $\left(d_f(0),d_{-f}(0)\right)$, the initial distances to the technology frontier, and $\left(a_f(t),a_{-f}(t)\right)_{t\geq 0}$, a pair of firms' strategy profiles, the performance measure of firm $f$ is
\begin{equation}\label{eqn:pmExt}
    \mathcal{J}_f\left(d_f(0),d_{-f}(0)\mid\left(a_f(t),a_{-f}(t)\right)_{t\geq 0}\right)
    = \mathds{E}\left\{\int_0^\infty e^{-\rho t}\left[\widetilde{\pi}_f\left(d_f(t),d_{-f}(t)\right) - \psi_f\left(a_f(t)\right)\right]dt\right\}. 
\end{equation} 
An equilibrium is a state-dependent strategy profile $$\left(a_A^*(t),a_B^*(t)\mid d_A(0),d_B(0)\right)_{t\geq 0}$$
such that given its rival's strategy $\left(a_{-f}^*(t)\right)_{t\geq 0}$, firm $f$'s strategy $\left(a_f^*(t)\mid d_A(0),d_B(0)\right)_{t\geq 0}$ maximizes its performance measure:
\begin{align}
    \left(a_f^*(t)\mid d_A(0),d_B(0)\right)_{t\geq 0} \in \argmax_{\left(a_f(t)\right)_{t\geq 0}} \mathcal{J}_f\left(d_f(0),d_{-f}(0)\mid\left(a_f(t),a_{-f}^*(t)\right)_{t\geq 0}\right).
\end{align}

Like in Section~\ref{sec:bslMdl}, we transform the model to a finite state--space equivalent and focus on its MPE, in which the state-dependent equilibrium value of firm $f$'s performance measure is characterized by its value function $v_f(d_f,d_{-f})$ and policy function $a_f^*(d_f,d_{-f})$. 

The argument for the existence of an MPE in this extended model is similar to that for the baseline model, only with the complication of having an extra state variable. Proposition~\ref{prp:existExt} expresses the system of equations for which the value functions $v_f$ and policy functions $a_f^*$ comprise a solution. The proof is skipped as it only requires a slight modification of the one for Proposition \ref{prp:exist}.

\begin{proposition}[MPE in the Extended Model]\label{prp:existExt}
    There exists an MPE $a_f^*(d_f,d_{-f})$ for $f\in\{A,B\}$, where $(d_f,d_{-f})\in\{0,\cdots,\ol{D}+\ol{m}\}^2$, $\min\{d_f,d_{-f}\}\leq \ol{D}$ and $|d_f-d_{-f}|\leq \ol{m}$. Moreover, the equilibrium strategy profile satisfies the following system of equations:
    \begin{align}
        v_f(d_f,d_{-f}) =& \dfrac{1}{\lambda\left[a_f^*(d_f,d_{-f}) + a_{-f}^*(d_{-f},d_{f})\right] + h\cdot\mathds{1}\{d_f\neq d_{-f}\} + \rho}
        \Big\{\widetilde{\pi}_f(d_f,d_{-f}) \nonumber \\
        &- \psi_f\left(a_f^*(d_f,d_{-f})\right) + \left[\lambda a_f^*(d_f,d_{-f}) + h\cdot\mathds{1}\{d_f > d_{-f}\}\right]v_f\left(T(d_f,d_{-f}\mid f)\right) \nonumber \\
        &+ \left[\lambda a_{-f}^*(d_{-f},d_f) + h\cdot\mathds{1}\{d_f < d_{-f}\}\right]v_f\left(T(d_f,d_{-f}\mid -f)\right)\Big\}, \label{eqn:existExtVal} \\
        \dfrac{d\psi_f\left(a_f^*(d_f,d_{-f})\right)}{da_f} 
        =& \lambda\left[v_f\left(T(d_f,d_{-f}\mid f)\right) - v_f(d_f,d_{-f})\right], \label{eqn:existExtPol} \\
        a_f^*(0,\ol{m})=&0. \label{eqn:existExtPolBnd}
    \end{align}
\end{proposition}

Due to the lack of an explicit expression of firm value function $v_{f}$, it is difficult to analytically characterize the impact of the shock. However, the following lemma shows that conditional on the technology gap, being one step further from the technology frontier always leads to lower values for both firms.

\begin{lemma}\label{lem:compValFcn}
    For $f\in\{A,B\}$ and $(d_f,d_{-f})\in\mathds{N}_+^2$, $$v_f(d_f,d_{-f})>v_f(d_f+1,d_{-f}+1)$$ 
    if and only if $v_f$ is defined at $(d_f,d_{-f})$ and $(d_f+1,d_{-f}+1)$.
\end{lemma}

Lemma~\ref{lem:compValFcn} implies that the leader's firm value decreases after being hit by a shock with degree one. However, this loss can be fully recovered by one innovation. In other words, innovation is now more valuable for the leader because it not only allows it to escape neck--and--neck competition with the laggard, but also helps it to recover the lost profit. As a corollary, firm $f$ has a higher value function if its rival lags further behind. 

\begin{corollary}\label{cor:compValFcn}
    For $f\in\{A,B\}$ and $(d_f,d_{-f})\in\mathds{N}_+^2$, $$v_f(d_f+1,d_{-f})<v_f(d_f,d_{-f})<v_f(d_f,d_{-f}+1)$$ if and only if $v_f$ is defined at all of these states.
\end{corollary}

%Directly from Lemma \ref{lem:compValFcn}, a negative profit shock lowers the values of both firms, and the severity of such a harm is increasing in the degree of shock $D$:
Proposition \ref{prp:inovShk} shows that the negative profit shock lowers the firm value, and that the magnitude of loss in firm value increases with the magnitude of the shock. 

\begin{proposition}\label{prp:inovShk}
    If there is a negative profit shock at time $t>0$, then for $f\in\{A,B\}$, $$v_f\left(d_f(t),d_{-f}(t)\right)<\displaystyle\lim_{s\rightarrow t^-}v_f\left(d_f(s),d_{-f}(s)\right),$$ and $\displaystyle\lim_{s\rightarrow t^-}v_f\left(d_f(s),d_{-f}(s)\right)-v_f\left(d_f(t),d_{-f}(t)\right)$ is strictly increasing in the degree of shock $D\in\{1,\cdots,\ol{D}\}$.
\end{proposition}
%%%%%%%%%%%%%%%%%%%%%%%%%%%%%%%%%%%%%%%%%%%%%%%%
\subsection{Responses to Negative Profit Shocks: Qualitative Statements}
The key theoretical results of the model concern firms' optimal R\&D effort in response to a negative profit shock. This subsection considers a shock with two extreme degrees of magnitude $D$. Proposition~\ref{prp:shkLarge} states that when $D$ is so large that both firms' post--shock profits are very small, they both reduce their R\&D efforts in response.
\begin{proposition}\label{prp:shkLarge}
    If $\delta>0$ is small enough, then there exists $\widehat{D}\in\{1,\cdots,\ol{D}\}$, such that if there is a negative profit shock with magnitude $D\geq\widehat{D}$ at time $t>0$ and $\min\left\{\displaystyle\lim_{s\rightarrow t^-}a_A^*(s),\displaystyle\lim_{s\rightarrow t^-}a_B^*(s)\right\}>0$, then $a_A^*(t)<\displaystyle\lim_{s\rightarrow t^-}a_A^*(s)$ and $a_B^*(t)<\displaystyle\lim_{s\rightarrow t^-}a_B^*(s)$.
\end{proposition}

% SC: revised YM003
% revision mark: YM003-JEDC
Intuitively speaking, when the destruction to profit is so overwhelming that it takes many successful innovations to recover, both firms are discouraged by the high R\&D costs for recovery. 
On the other hand, 
% revision mark: YM003-JEDC
when the magnitude of the shock is minimal, the leader is only one step away from the technology frontier. Compared to the pre--shock state, the leader has a greater incentive for R\&D : it not only enhances the leader's technological  superiority, but also recovers the loss from the negative profit shock. This intuition is formalized in the following proposition.

\begin{proposition}\label{prp:shkSmall}
    If a negative profit shock with $D=1$ occurs at time $t>0$ and $\displaystyle\lim_{s\rightarrow t^-}d_f(s)=0$, then $a_f^*(t)>\displaystyle\lim_{s\rightarrow t^-}a_f^*(s)$. 
\end{proposition}

Propositions~\ref{prp:shkLarge} and \ref{prp:shkSmall} indicate that when $\delta$ is small enough, there exists a threshold of the shock's magnitude, beyond which firm $f$ responds by reducing R\&D effort. Denote this threshold by $\widehat{D}_f(d_f,d_{-f})$ to reflect the fact that it depends on the state just prior to the shock. For example, if firm $f$ is on the technology frontier prior to the shock, then $\widehat{D}_f(0,d_{-f})$ is strictly greater than $1$.

Suppose firm $A$ is the leader before the shock. Following the above analysis, if $\widehat{D}_A(0,d_B)>\widehat{D}_B(d_B,0)$, then there exists a degree of shock $D$ satisfying $\widehat{D}_B(d_B,0)\leq D<\widehat{D}_A(0,d_B)$ such that firm $A$ responds by higher R\&D effort while firm $B$ responds by lower R\&D. As a consequence, for some post--shock periods, the expected technology gap is strictly higher than the pre--shock level. 
%%%%%%%%%%%%%%%%%%%%%%%%%%%%%%%%%%%%%%%%%%%%%%%%
%%%%%%%%%%%%%%%%%%%%%%%%%%%%%%%%%%%%%%%%%%%%%%%%
\section{Data and Negative Profit Shocks Construction}\label{sec:empirics}
The model is informative of how firms react to a negative profit shock, but these are qualitative statements that depend on the magnitude of the shocks. In this section, we describe the various datasets we use and our construction of the negative profit shocks. 
\subsection*{Data Sources}

The CRSP/Compustat Merged (CCM) Database provides information about firm fundamentals from 1957 to 2011. The database contains a subset of U.S. publicly listed firms that can be identified in two ways: (1) CRSP’s permanent company and security identifiers; (2) Compustat’s permanent company identifier. We use the following variables: R\&D expense, sale, number of employees, book value of capital and value added output.

The NBER-CES Manufacturing Database allows us to obtain the industry-level cost shares and prices deflators from 1958 to 2011. It provides information about the labor share of output and contains variables like industry-level payroll, value added, shipment price deflator and investment price deflator. Combining these variables with the firm fundamentals, we construct revenue TFP (TFPR), which is the measure of productivity that we use in Section~\ref{sec:emp}. 

To evaluate the outcome of firms' R\&D activities, we employ the patent value dataset made available by \cite{KPSS2017}. The dataset contains the private value (or market value) and scientific value of each U.S. patent issued by the USPTO from 1926 to 2010 that can be matched to a publicly listed firm. \cite{KPSS2017} measure a patent's economic value by the stock market response to the news of the issuance within a two--day window. Aggregating patent-level values to firm-year level, the authors match the outcomes to the CCM database using permanent company and security identifiers.

The patent dataset contains two main variables: firm-year private patent value and firm-year scientific patent value. Both measures are scaled by firm size approximated by book assets. The difference is that patents in the latter category are weighted by the number of forward citations. In the main text we use the private value of patents, but robustness checks show that our empirical findings hold for scientific values as well. \footnote{\cite{KPSS2017} also provide levels of innovation values, but these unscaled measures are sensitive to firm size, which makes it difficult to compare and interpret innovations done by firms with different ages. Therefore we do not use them for this paper.}

We merge the three databases by the four-digit Standard Industrial Classification (SIC) codes and obtain an unbalanced annual panel from 1970 to 2010, covering 4,074 U.S. manufacturing firms from 135 industries. Details of data pre-processing and variable constructions are available in Appendix \ref{asc:data}.

\subsection*{An Index of Negative Profit Shocks}
In this paper, a negative profit shock refers to the event for which an industry's aggregate annual profit falls below a certain threshold. We first scale each firm's annual gross profit by its total assets 
% revision mark: YM001-JEDC
so that industries with different ranges of firm size become comparable.\footnote{According to the U.S. Generally Accepted Accounting Principles (GAAP), a firm's gross profit is defined as the difference between sales and cost of goods sold.} The scaled gross profit is denoted as $gp_{ijt}$, in which subscripts $i$, $j$ and $t$ represent firm, industry and year respectively. We run the following regression to detrend the variable:

\begin{equation}\label{eqn:detrend}
    gp_{ijt} = \alpha + \delta_t + \delta_j\times t + u_{ijt},
\end{equation}
where $\delta_t$ and $\delta_j$ denote the year and industry fixed effects. The residual, $gp^\text{res}_{ijt}$, is orthogonal to both the year fixed effect and industry--specific linear trend, and we treat it as the detrended firm--year scaled profit.

Let $gp_{jt}^\text{res}$ denote the aggregation of $gp^\text{res}_{ijt}$ to the industry--year level. We compute the 5th percentile of $gp_{jt}^\text{res}$ over all industry--year cells and denote it as $\text{pct}_5(gp^\text{res})$. A negative profit shock is defined as the event for which an industry-year observation of $gp_{jt}^\text{res}$ falls below $\text{pct}_5(gp^\text{res})$. More specifically, we define a dummy variable $NPS_{jt}$ for negative profit shock as follows:
\begin{equation}\label{eqn:npsDef}
    NPS_{jt}
    =
    \begin{cases}
        1   & \quad \text{if} \ \displaystyle gp_{jt}^\text{res}<\text{pct}_5(gp^\text{res}); 
        \\
        0   & \quad \text{otherwise}.
    \end{cases}
\end{equation}
By construction, $NPS_{jt}$ takes value one if and only if there is a negative profit shock on the corresponding industry-year cell. Note that throughout the construction we stay agnostic about the source of the shocks and only require that they are exogenous to the firms. We also consider using 1st and 10th percentiles as thresholds, and they have economic interpretations: the smaller the percentile, the more drastic the shock.
%%%%%%%%%%%%%%%%%%%%%%%%%%%%%%%%%%%%%%%%%%%%%%%%
%%%%%%%%%%%%%%%%%%%%%%%%%%%%%%%%%%%%%%%%%%%%%%%%
\section{Quantitative Analysis}\label{sec:Quant}
The theoretical analysis in Sections \ref{sec:bslMdl} and \ref{sec:extMdl} is qualitative. We now take the model to the data and quantitatively illustrate the key mechanism: firms' heterogeneous responses to the shock induce a larger productivity dispersion. We embed the model in a general equilibrium framework in Section~\ref{sec:parMdl}, describe calibration and estimation in Section~\ref{sec:cali and est}, numerically analyze the model in Sections~\ref{baseline numerical} and \ref{extend numerical}, and present firms' responses in a simulated economy in Section~\ref{sec:sim}.
%%%%%%%%%%%%%%%%%%%%%%%%%%%%%%%%%%%%%%%%%%%%%%%%
\subsection{Model Parameterization}\label{sec:parMdl}
\textit{Consumers.} --- Following \cite{AHHV2001}, we assume there is a unit mass of households who consume goods from a continuum of industries indexed by $\omega\in[0,1]$. The household utility function is
\begin{equation}\label{eqn:utlFcn}
    U = \displaystyle \int_0^\infty e^{-\rho t}\left[\int_0^1\log x_\omega(t)d\omega -L(t)\right]dt,
\end{equation} 
where $x_\omega(t)$ denotes consumption at time $t$ of industry $\omega$ output, $L(t)$ denotes the labor supply and $\rho>0$ denotes the rate of time preference. Each industry $\omega$ has two firms, who produce outputs $x_{\omega,A}$ and $x_{\omega,B}$ respectively. The industry output $x_\omega$ satisfies
\begin{equation}\label{eqn:aggProd}
    x_\omega = \left(x_{\omega,A}^\alpha + x_{\omega,B}^\alpha\right)^{1/\alpha}, \quad \alpha\in[0,1],
\end{equation}
where the parameter $\alpha$ controls the degree of substitutability between the two goods, and is constant across industries. 

By the log utility assumption in equation~(\ref{eqn:utlFcn}), in equilibrium consumers spend equal amount over each industry. This common amount is normalized to unity by using expenditure as the numeraire for the prices $p_{\omega,A}$ and $p_{\omega,B}$. The representative consumer thus chooses $x_{\omega, A}$ and $x_{\omega, B}$ to maximize $x_\omega$, subject to the budget constraint $p_{\omega,A}x_{\omega,A}+p_{\omega,B}x_{\omega,B}=1$.
\vskip 0.3cm
\textit{Producers.} --- Firm $f$ uses labor as the unique input and takes the unit wage rage as given. We consider the following production function:
\begin{equation}\label{eqn:prodFcn}
    y_f = \gamma^{n_f} L_f, \quad \gamma>1,
\end{equation}
where $L_f$ denotes the quantity of labor hired by firm $f$. Its productivity, $\gamma^{n_f}$, is strictly increasing in its technology level $n_f$, which is modeled as its position on the technology ladder in Section \ref{sec:bslMdl}.

By Proposition 1 in \cite{AHHV2001}, for $\alpha\in(0,1)$, firm $f$'s profit is jointly determined by the following two equations: 
\begin{align}
    \pi(\Delta n_f\mid\alpha,\gamma) = & \dfrac{\zeta_f(\Delta n_f \mid \alpha,\gamma)(1-\alpha)}{1-\alpha\zeta_f(\Delta n_f \mid \alpha,\gamma)}; \label{eqn:piSpec1} \\
    (1-\alpha\zeta_f)^\alpha \zeta_f\gamma^{-\alpha\Delta n_f} =& \left(1-\alpha(1-\zeta_f)\right)^\alpha(1-\zeta_f), \label{eqn:piSpec2}
\end{align}
where $\zeta_f$ is firm $f$'s revenue. The profit function satisfies Assumption \ref{asm:pi}. Following the empirical literature on estimating innovation cost functions \citep{AkcigitAtes2021}, we assume the R\&D cost function is quadratic:
\begin{equation}\label{eqn:psiSpec}
    \psi_f(a_f) = \dfrac{\kappa_f}{2}a_f^2,
\end{equation}
where $\kappa_{f}$ denotes firm $f$'s cost coefficient. The quadratic form satisfies the regularity conditions in Assumption~\ref{asm:psi}. Without loss of generality, we treat firm $A$ as the low--cost firm. 
%%%%%%%%%%%%%%%%%%%%%%%%%%%%%%%%%%%%%
\subsection{Calibration and Estimation}\label{sec:cali and est}
We use the baseline model for structural estimation because the only extra parameter in the extended model is $\delta$, the severity of a negative profit shock with a unit magnitude. The choice of its value does not affect the other parameters because the shock is exogenous. 

\subsubsection*{Calibration}

Following \cite{AHHV2001}, we set the discount rate at $\rho=0.03$, which is the labor rate of interest since the numeraire is consumer expenditure and the wage rate is one. The multiplier on the R\&D effort is normalized to one.

% SC m=2 paragraph
Now we turn to the choice of the maximum technology gap $\ol{m}$. Unless the laggard can easily imitate the leader, the technology gap has a high probability mass close to its maximum value in the pre-shock equilibrium. Therefore, the post-shock percentage change in the technology gap is decreasing in $\overline{m}$. Because there is no empirics to discipline the choice of $\overline{m}$, we set it to be two.

%The maximum technology gap $\ol{m}$ is set to two. A smaller number $\ol{m}=1$ is insufficient to generate the cross--firm gap of variables observed in the data. A larger $\ol{m}$, on the other hand, cannot generate a large enough response of the technology gap to the shock because the gap is too close to the maximum value in the pre-shock stationary equilibrium\footnote{\textcolor{blue}{
% revision mark: YM002-JEDC
%When the imitation rate is not large enough, the technology gap has a high probability mass close to its maximum value in the pre-shock equilibrium. Therefore, the post-shock percentage change in the technology gap is decreasing in $\overline{m}$. Given that no empirical observation justifies another choice of $\overline{m}$, we set $\overline{m} = 2$. 
%}}. 

To calibrate the R\&D cost coefficients $\kappa_{A}$ and $\kappa_{B}$, we first combine equations~(\ref{eqn:aFoc}) and (\ref{eqn:psiSpec}) to obtain the analytical expression from the model:
\begin{equation}\label{eqn:caliKappa} 
    \kappa_f = \dfrac{\left[v_f(\Delta n_f + 1)-v_f(\Delta n_f)\right]^2}{2\psi_f\left(a_f^*(\Delta n_f)\right)}, \qquad \forall \ \Delta n_f,
\end{equation}
where $v_f(\Delta n_f + 1)-v_f(\Delta n_f)$ represents the gain in firm value from a new innovation. Since the added value of an innovation to a firm can be reflected by stock market responses to news of patent issuance, we use the firm--year value of patents, scaled by sale, as a proxy.

The R\&D cost function $\psi_{f}(\cdot)$ in the denominator is unobservable, but Lemma~\ref{lem:rdi} states that the low--cost firm on average invests more in R\&D than the high--cost firm. Therefore, we use firm--year log RDI as a proxy for the equilibrium R\&D cost $\psi_f(a^*)$. 

Firms' RDI from different industries are not directly comparable because the threshold for ``intensive'' spending varies across industries. Therefore, we first rank firms in the same industry using the within--industry \textit{empirical cumulative distribution function} (ECDF) of log RDI. More specifically, the ECDF of firm $i$ in industry $j$ at year $t$ is defined as 

\begin{equation*}\label{eqn:ecdfRdi}
    ECDF_{ijt} = \frac{1}{n_{jt}}\sum_{i^\prime} \mathds{1}\{\log(RDI_{i^\prime jt})< \log(RDI_{ijt})\},
\end{equation*} 
where $n_{jt}$ denotes the total number of firms in industry $j$ at year $t$. We then sort firms in each industry-year cell into ten groups based on their ECDF ranks:
\begin{equation}\label{eqn:decileDef}
    \ol{c}_{jt}^s = \argmax_{i^\prime}\left\{\log(RDI_{i^\prime jt}):ECDF_{i^\prime jt}\leq s/10\right\}, \quad s=1,\cdots,10.
\end{equation}
Then we create ten dummy variables that denote the decile group to which each firm belongs:
\begin{align}
    c_{ijt}^1 =& 
    \begin{cases} 
        1 & \ \text{if} \quad \log(RDI_{ijt})\leq \ol{c}_{jt}^1, \\
        0 & \ \text{otherwise};  
    \end{cases} \label{eqn:regGroup1} 
    \\
    c_{ijt}^s =& 
    \begin{cases}
        1 & \ \text{if} \quad \ol{c}_{jt}^{s-1} < \log(RDI_{ijt})\leq \ol{c}_{jt}^s \ \text{and} \ 2\leq s \leq 9,\\
        0 & \ \text{otherwise};  
    \end{cases} \label{eqn:regGroup2} 
    \\  
    c_{ijt}^{10} =& 
    \begin{cases}
        1 & \ \text{if} \quad \log(RDI_{ijt})> \ol{c}_{jt}^9, \\
        0 & \ \text{otherwise}.  
    \end{cases} \label{eqn:regGroup3}  
\end{align}
For example, $c_{ijt}^3=1$ means that firm $i$'s log RDI at year $t$ falls between the second and third deciles with respect to all firms in the same industry-year cell. 

We treat the firms with $c_{ijt}^3=1$ as representatives for the low-cost firm in our model, and those with $c_{ijt}^8=1$ for the high-cost firm. The choice corresponds to the definition of IQR. As such, we calibrate $\kappa_f$ as follows:
\begin{align}
    \kappa_A =& \dfrac{1}{JT}\sum_{j,t}\dfrac{1}{n_{jt}}\Big[\sum_{i}
    \dfrac{(Tsm_{ijt})^2/2}
    {rdi_{ijt}}\times \mathds{1}\{c_{ijt}^8=1\}\Big]; \label{app:caliKappaA} \\
    \kappa_B =& \dfrac{1}{JT}\sum_{j,t}\dfrac{1}{n_{jt}}\Big[\sum_{i}
    \dfrac{(Tsm_{ijt})^2/2}
    {rdi_{ijt}}\times \mathds{1}\{c_{ijt}^3=1\}\Big], \label{app:caliKappaB}
\end{align}
where $Tsm_{ijt}$ denotes the firm-year innovation value, $rdi_{ijt}$ denotes the RDI, $J$ and $T$ are the numbers of industries and years respectively, and $n_{jt}$ is the number of firms in industry-year cell $(j,t)$. Indicators $\mathds{1}\{c_{ijt}^8=1\}$ and $\mathds{1}\{c_{ijt}^3=1\}$ are defined in equation~(\ref{eqn:regGroup2}). Table~(\ref{tbl:calibration}) summarizes the calibrated parameters. 
\begin{table}[ht]
    \centering
    \begin{threeparttable}
        \caption{Calibrated Parameters} \label{tbl:calibration}
        \begin{tabular}{cllc}
        \hline\hline
        \textbf{Parameter}   & \textbf{Description}                        & \textbf{Value}          \Tstrut\Bstrut\\
        \hline
        $\rho$      & discount factor                    & $0.03$        \Tstrut\Bstrut\\
        $\lambda$   & multiplier on R\&D effort          & $1$            \Tstrut\Bstrut\\
        $\ol{m}$    & maximum technology gap             & $2$            \Tstrut\Bstrut\\
        $\kappa_A$  & R\&D cost parameter of firm $A$    & $0.0167$       \Tstrut\Bstrut\\
        $\kappa_B$  & R\&D cost parameter of firm $B$    & $0.0259$       \Tstrut\Bstrut\\
        \bottomrule
        \end{tabular}
    \end{threeparttable}    
\end{table}

\subsubsection*{GMM Estimation}
The remaining three parameters to be determined are elasticity of substitution $\alpha$, the parameter in the production function $\gamma$, and the imitation rate $h$. These parameters do not feature analytical moment conditions like $\kappa_{f}$, but the analytical form of the stationary distribution in Proposition \ref{prp:statDist} allows us to estimate them by the \textit{generalized method of moments} (GMM) method. We choose three target moments that have analytical expressions in the model: the expected ratio of the RDI of low-cost firms to high-cost ones, the expected ratio of firm values, and the expected ratio of the gross profits. The estimation details are available in Appendix~\ref{Computation Appendix: GMM}.

\begin{table}[!htbp]
    \centering
    \begin{threeparttable}
        \caption{Parameter Estimates} \label{tbl:para}
        \begin{tabular}{cllc}
        \hline\hline
        \textbf{Parameter}   & \textbf{Description}                                   & \textbf{Estimate}           & \textbf{Standard Errors}         \Tstrut\Bstrut\\
        \hline
        $\alpha$    & Elasticity of substitution       & $0.9936$               & (0.013)      \Tstrut\Bstrut \\
        $\gamma$    & Parameter in the production function          & $1.0286$              & (0.017)      \Tstrut\Bstrut \\        
        $h$         & Imitation rate                                & $0.4041$                & (0.010)      \Tstrut\Bstrut\\
        \bottomrule
        \end{tabular}
    \end{threeparttable}    
\end{table}

% In the revision, redo the quantitative part and hence rewrite everything. 

Table (\ref{tbl:para}) tabulates the estimated parameters. The estimated value of $\alpha$ is 
{\textcolor{blue}{
% revision mark: YM
close to
}}
$1$, which suggests 
{\textcolor{blue}{
% revision mark: YM
nearly
}}
perfect substitutability between the two products, and thus intense competition in the product market.\footnote{
{\textcolor{blue}{
% revision mark: YM005-JEDC
This is not surprising given the model's assumption that there are infinitely many industries in the economy.
}}
} These parameters generate some non--targeted moments that are overall close to those in the data, as reported in Table~(\ref{tbl:fit}). One exception is the average relative market value as the the model overvalues the low-cost firm relative to the high-cost firm. One possible explanation is that, in our model the only cross-firm heterogeneity is the R\&D capacity, but in reality there are many other dimensions that explain the variation in firms' market values.

\begin{table}[ht]
    \centering
    \begin{threeparttable}
        \caption{Model and Data Moments (Non--Targeted)} \label{tbl:fit}
        \begin{tabular}{lcc}
        \hline\hline
        \textbf{Moments}                            & \textbf{Data}      & \textbf{Model}      \Tstrut\Bstrut\\
        \hline
        First order condition for low-cost firms                               & $0.0167$  & $0.0167$   \Tstrut\Bstrut\\
        First order condition for high-cost firms                              & $0.0259$  & $0.0259$  \Tstrut\Bstrut \\
        Average relative R\&D intensity, low-cost to high-cost firms           & $24.8303$ & $21.4103$  \Tstrut\Bstrut\\
        Average relative market value                                          & $12.5281$ & $19.9372$  \Tstrut\Bstrut\\ 
        Average relative gross profit                                          & $9.2506$  & $8.4235$   \Tstrut\Bstrut\\     
        \bottomrule
        \end{tabular}
    \end{threeparttable}    
\end{table}
%%%%%%%%%%%%%%%%%%%%%%%%%%%%%%%%%%%%%%%%%%%%%%%%
%%%%%%%%%%%%%%%%%%%%%%%%%%%%%%%%%%%%%%%%%%%%%%%%
\subsection{Numerical Analysis of the Baseline Model} \label{baseline numerical}
%With the parameterized model, we continue to study its properties in the stationarity equilibrium, and how it respond to the negative profit shock, especially whether such responses are qualitatively and quantitatively similar to our empirical findings in Section \ref{sec:irf}.

%We firstly solve the baseline model numerically to obtain the value functions and policy functions of the two firms. Then we turn to the quantitative study of the extended model, where the state is two-dimensional in firms' distances to the technology frontier. The simulation of the extended model shows firms's heterogeneous responses in R\&D to the NPS, and helps to explain the countercyclical productivity dispersion.

We numerically solve the baseline model using a value function iteration procedure that is described in detail in Appendix \ref{Computation Appendix: VFI}. Figure (\ref{fig:solBsl}) plots the firms' value functions and policy functions. The left panel conveys two messages. First, both firms' value functions are strictly increasing in the technology gap. This is intuitive as a larger technology gap leads to a higher profit flow. Second, the low-cost firm's value strictly dominates that of the high-cost firm's in each state. This is consistent with Lemma \ref{lem:valFcn}, which states that in equilibrium, the low-cost firm achieves a performance measure at least as high as that of the high-cost firm.

\begin{figure}[!htbp]
  \centering
  \caption{Numerical solution of the baseline model}\label{fig:solBsl}
  \includegraphics[width=1\textwidth]{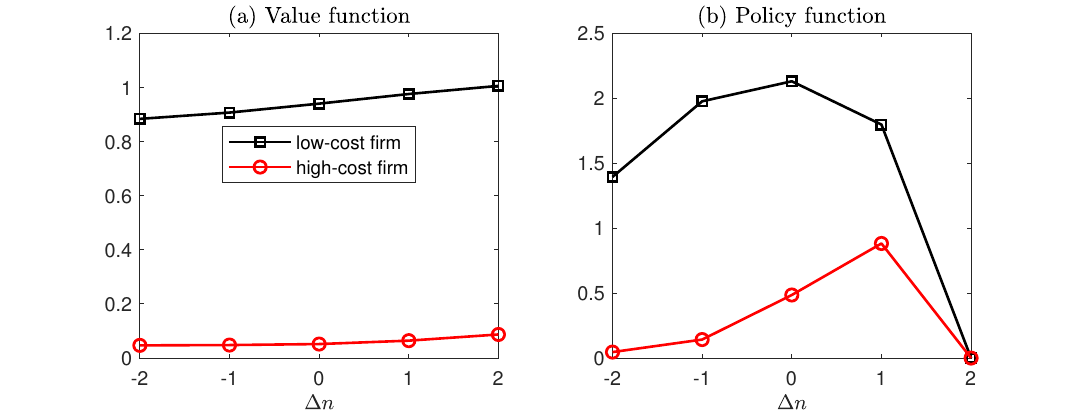}
  \begin{minipage}{0.8\textwidth}
	{\footnotesize Note: The horizontal axis denotes the technology gap. In panel (b), each firms exerts zero effort if they are at the maximum technology gap. This is due to the automatic catch--up assumption. \par}
  \end{minipage}
\end{figure}

The right panel of Figure~(\ref{fig:solBsl}) shows that the highest effort from the low--cost firm happens when it is neck--and--neck with the high--cost firm $(\Delta n = 0)$. The low--cost firm is incentivized to be more intensive in R\&D so as to escape the competition. On the other hand, the high--cost firm exerts the highest R\&D effort when it is one step ahead. One explanation is that the high--cost firm anticipates its low--cost rival's aggressiveness at gap zero and hence chooses not to exert the highest effort; on the other hand, when it is ahead, it exerts the greatest effort in the hopes of solidifying its lead. Both firms exert zero effort at the maximum gap due to the automatic catch--up assumption. 

The right panel of Figure~(\ref{fig:solBsl}) also confirms that Lemma \ref{lem:rdi} extends to cases in which $\overline{m}>1$, namely when the R\&D effort of the low-cost firm strictly dominates that of the high-cost one except at the maximum gap. Consequently, firm $A$ is the leader under expectation in the stationary equilibrium.

\begin{figure}[ht]
  \centering
  \caption{Stationary Distribution of Technology Gap}\label{fig:limDist}
  \includegraphics[]{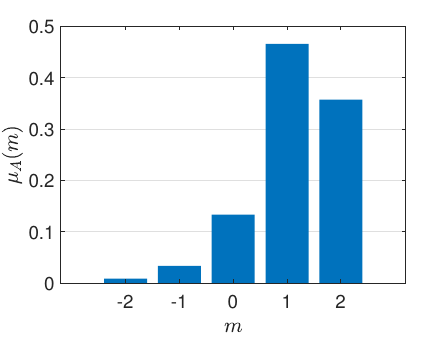}
  \begin{minipage}{0.8\textwidth}
	{\footnotesize Note: The horizontal axis denotes the technology gap. \par}
  \end{minipage}
\end{figure}

Figure (\ref{fig:limDist}) plots the stationary distribution of the technology gap from firm $A$'s point of view. For any $m>0$, the stationary probability for firm $A$'s technology gap to be $\Delta n_A=m$ is greater than that of $\Delta n_A=-m$. As $t\rightarrow \infty$, firm $A$ has an 82.34\% (that is, $\mu_A(1)+\mu_A(2)$) chance being the leader, a $4.31\%$ chance of being the follower, and a $13.35\%$ chance of being neck--and--neck with firm $B$. Thus firm $A$ is expected to be the leader in the long run, with an expected gap of $1.13$. 

%%%%%%%%%%%%%%%%%%%%%%%%%%%%%%%%%%%%%%%%%%%%%%%%
\subsection{Numerical Analysis of the Extended Model} \label{extend numerical}
In the extended model, the profit function takes the form (\ref{eqn:piExt}). Because it is empirically hard to disentangle $D$ and $\delta$, we set $\delta = 0.05$, which means a shock of degree one reduces the profit of each firm by $5$ percent. The magnitude of the shock is manifested via $\min\{d_{A}, d_{B}\}$. The value function $v_f(d_f,d_{-f})$ and policy function $a_f^*(d_f,d_{-f})$ are again solved numerically using value function iteration.

%This makes the profit function depends on both $d_A$ and $d_B$ -- the distances of the two firms to the technology frontier -- as do the value function and policy function in the extended model.  

\begin{figure}[ht]
  \centering
  \caption{Policy functions in the extended model}\label{fig:solExt}
  \includegraphics[width=1\textwidth]{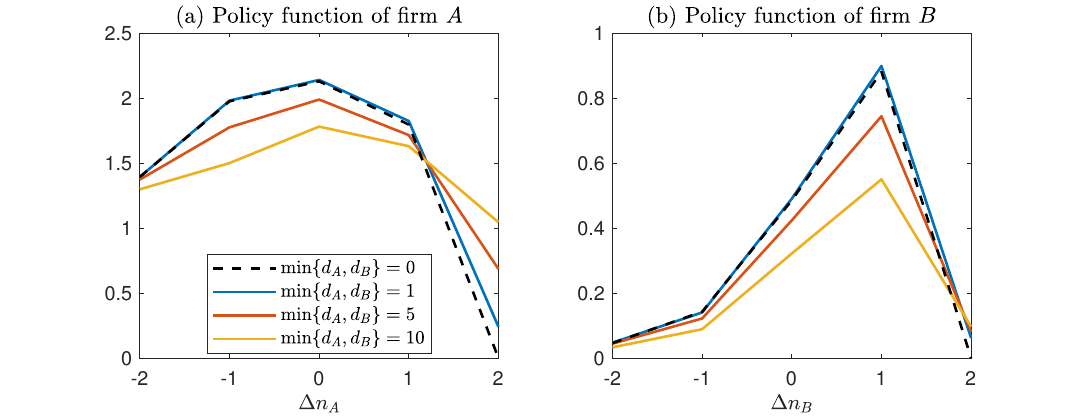}
  \begin{minipage}{0.8\textwidth}
      {\footnotesize Note: The horizontal axis represents each firm's technology gap, and different curves plot the firm's policy functions under different degrees of shock. \par}
  \end{minipage}
\end{figure}

Figure~(\ref{fig:solExt}) plots firms' policy functions for shocks with varying magnitudes. When firm $f$ is the leader ($\Delta n_f>0$), it responds to a shock of degree one by increasing its R\&D effort, as predicted by Proposition \ref{prp:shkSmall}. For shocks with larger magnitudes, however, both firms reduce R\&D efforts as long as they are not at the maximum gap, which is consistent with Proposition \ref{prp:shkLarge}. If a firm is at the maximum gap when the shock happens, it will impose greater R\&D effort regardless of the shock magnitudes due to the strong incentive to recover the lost profit.

%For the mechanism to work, we require that the leader needs to be at the maximum gap and a shock with medium or large magnitudes. 

%Given the policy function in Figure~(\ref{fig:solExt}), if there is a substantial chance that a firm is at the maximum gap, then on average, a shock with a medium or large degree will raise its R\&D effort and reduce that of its rival's. This is the mechanism explaining larger gap in R\&D effort and thus in productivity as found empirically in Section \ref{sec:irf}. In the following subsection, we produce the impulse response functions by simulation to further verify the proposed mechanism.
%%%%%%%%%%%%%%%%%%%%%%%%%%%%%%%%%%%%%%%%%%%%%%%%
\subsection{Simulating the Impact of a Negative Profit Shock}\label{sec:sim}
The economy starts with firms being neck--and--neck on the frontier and reaches the stationary distribution before the shock occurs. We hit the economy with a shock of magnitude $D=4$. Conditional on $\delta=0.05$ in equation~(\ref{eqn:piExt}), this shock lowers the industry's profit by about $20\%$.\footnote{In Section~\ref{appendix:estNPS}, we regress aggregate profits of firms on the negative profit shocks, and column one of Table~\ref{tbl:effNPS} shows that the impact is -0.192.} Details of simulation procedures are available in Appendix~\ref{Computation Appendix: Simulations}. 

\begin{figure}[ht]
  \centering
  \caption{Simulation of the impact of negative profit shock}\label{fig:sim}
  \includegraphics[width=1\textwidth]{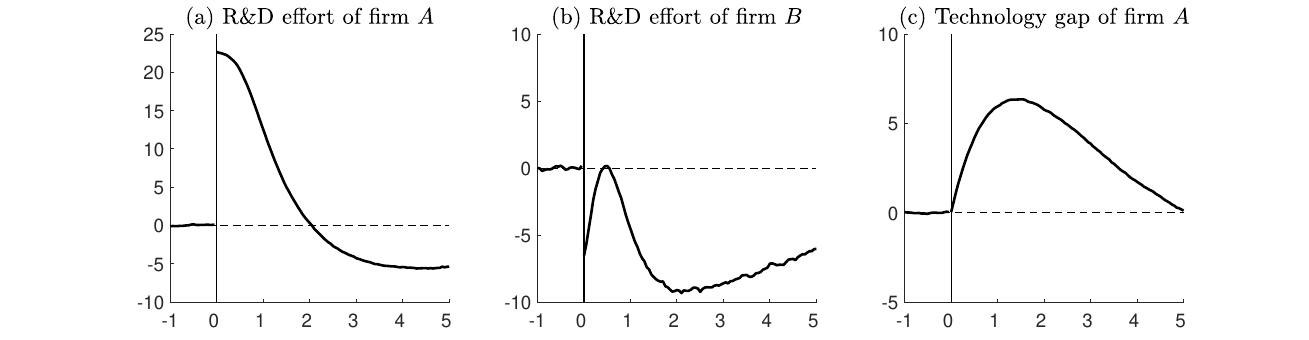}
  \begin{minipage}{0.8\textwidth}
	{\footnotesize Note: The three panels show percentage deviations of variables from their respective stationary levels. One unit along the horizontal axis corresponds to one year. Time $0$ is the period when the negative profit shock hits. \par}
  \end{minipage}
\end{figure}

Figure~(\ref{fig:sim}) plots the simulated impulse responses of firms' efforts and technology gaps from the low--cost firm's point of view. The straight lines imply that the economy is in stationary equilibrium before the shock. At the onset the shock at time 0, the low-cost firm $A$ instantly increases its R\&D effort, and the high-cost firm $B$ responds by reducing its R\&D effort. After the shock, firm $B$ faces two opposing incentives for R\&D. On one hand, it can gain more from R\&D as the leader innovates and regains industrial--level profitability. On the other hand, it is discouraged from R\&D because the leader's innovation enlarges the technology gap. As the technology enlarges over time, the second incentive dominates, and hence we observe the shape as shown in panel (b). % YM007-JEDC
In the meantime, the increasing gap also reduces the leader's R\&D incentive, so its R\&D effort drops below the pre-shock level after an initial jump, as shown in panel (a).

Over time, the difference in R\&D efforts is higher than the pre--shock level. These movements in opposite directions cause a hump-shaped increase in the technology gap, as shown in panel (c). This change in technology gap is gradual because it takes time for the effect of an instant change in the Poisson arrival rate to be seen. In the longer horizon, as the the difference in R\&D converges back to the pre--shock level, the technology gap closes.  

As shown in panel (c), the simulated impulse response functions reaches $6.35\%$ at its peak. To understand how much our model can potential explain the observed countercyclical movements, we regress the industry-year IQR of log TFP on the negative profit shocks. The second column of Table~(\ref{tbl:effNPS}) in Section~\ref{appendix:estNPS} shows that the shock is associated with a $9.56\%$ increase in productivity dispersion. Therefore, our model can explain two-thirds of the movement. 
%In the next section , we present further empirical evidence of our theory on countercyclical productivity dispersion, with reduced-form estimations.
%%%%%%%%%%%%%%%%%%%%%%%%%%%%%%%%%%%%%%%%%%%%%%%%%%
%%%%%%%%%%%%%%%%%%%%%%%%%%%%%%%%%%%%%%%%%%%%%%%%%%
\section{Empirical Evidence of Heterogeneous Responses} \label{sec:emp}
The quantitative model predicts that in response to a negative profit shock of realistic magnitude, the low-cost firm increases R\&D effort while the high-cost firm does the opposite. This section provides supportive evidence by testing the following model implication: after the shock, firms that previously had high RDI increase RDI, while firms that previously had low RDI do the opposite. 

%,,We consider two approaches: (i) ECDF and (ii) sorting that allows for non--linear responses. Both measures are constructed in Section~\ref{sec:Quant}. 
%%%%%%%%%%%%%%%%%%%%%%%%%%%%%%%%%%%%%%%%%%%%%%%%%%
\subsection{ECDF Approach}
Recall that we use the firm--year RDI as a proxy for the equilibrium R\&D cost, and we use the within--industry ECDF to address the challenge of incompatibility of firms' RDI standards across different industries. We consider the following regression:
\begin{equation}\label{eqn:regSpec1}
    \log(RDI_{ijt}) = \beta NPS_{j,t-1} + \gamma NPS_{j,t-1}\times ECDF_{i,j,t-1} + \eta ECDF_{i,j,t-1} + \boldsymbol{\lambda}\boldsymbol{X}_{ijt} + \alpha_i + \delta_t + u_{ijt},
\end{equation}
where $NPS$ denotes the negative profit shock, $\alpha_i$ and $\delta_t$ denote firm and year fixed effects, and the set of control variables $\boldsymbol{X}$ includes the log of TFP, capital and employment, market share, number of firms in the industry-year cell, and the Herfindahl-Hirschman Index (HHI).\footnote{
% revision mark: YM006-JEDC
Although the model implies a strong competition in a continuum of highly segmented markets, we impose no restriction on the market structure in our reduced-form analysis. Still, we include the HHI as a covariate to ensure that the variation in productivity dispersion is not driven by changes in with-industry competitiveness.}
We use log RDI instead of level because the latter is highly right-skewed, with mean $4.49$ and the maximum as high as $25684.40$. This may be due to the fact some firms may have sales close to zero at times when their R\&D expenditures are positive. The coefficient $\gamma$ captures the heterogeneity of firms' responses. A positive estimate means that compared to firms with low past RDI, those with higher past--year RDI respond to the shock by increasing RDI.

One potential concern, however, is that a high annual fluctuation in RDI can make the previous--year ECDF a bad indicator of a firm's long--term capacity to do R\&D. To smooth the fluctuation, we use the three-year moving average of R\&D expenditure to construct RDI, and compute the corresponding ECDF, denoted as $ECDF^{MA3}$. For robustness checks, we run the same specification for the three--year moving averages. 

Table (\ref{tbl:heterResp}) tabulates the regression results. The first two columns use $ECDF$ and the other two use $ECDF^{MA3}$. The first row reports the average impact of the negative profit shock on the RDI of the least R\&D intensive firms, whose ECDF values of past RDI are close to zero. This impact is negative for all four specifications. The standard errors are clustered at industry level since the negative profit shocks affect all firms in an industry. We apply the method by \cite{LZ1986}, with the caveat that the resultant standard errors might be conservative \citep{AAIW2017}. 
\begin{table}[!ht]
    \centering
    \begin{threeparttable}
        \caption{Responses of R\&D intensity to the negative profit shock} \label{tbl:heterResp}
        \begin{tabular}{lcccccccc}
        \hline\hline
        $\log(RDI_t)$     & \quad         & (1)           & \qquad        & (2)           & \qquad        & (3)           & \qquad        & (4)         \Tstrut\Bstrut   \\
        \hline
        $NPS_{t-1}$       &               & $-0.188^{**}$ &               & $-0.212^{***}$&               & $-0.300^{**}$ &               & $-0.385^{**}$  \Tstrut\Bstrut\\
                          &               & $(0.074)$     &               & $(0.061)$     &               & $(0.138)$     &               & $(0.157)$     \Tstrut\Bstrut \\
        $NPS_{t-1}\times ECDF_{t-1}$
                          &               & $0.639^{***}$ &               & $0.521^{***}$ &               &               &               &               \Tstrut\Bstrut \\
                          &               & $(0.179)$     &               & $(0.113)$     &               &               &               &                \Tstrut\Bstrut\\
        $ECDF_{t-1}$
                          &               & $1.338^{***}$ &               & $1.244^{***}$ &               &               &               &               \Tstrut\Bstrut \\
                          &               & $(0.127)$     &               & $(0.094)$     &               &               &               &               \Tstrut\Bstrut \\
        $NPS_{t-1}\times ECDF_{t-1}^\text{MA3}$
                          &               &               &               &               &               & $0.790^{**}$  &               & $0.785^{**}$  \Tstrut\Bstrut \\
                          &               &               &               &               &               & $(0.305)$     &               & $(0.289)$     \Tstrut\Bstrut \\
        $ECDF_{t-1}^\text{MA3}$
                          &               &               &               &               &               & $1.083^{***}$ &               & $1.084^{***}$ \Tstrut\Bstrut \\
                          &               &               &               &               &               & $(0.154)$     &               & $(0.149)$      \Tstrut\Bstrut\\                                        
                          &               &               &               &               &               &               &               &               \Tstrut\Bstrut \\
        Control variables &               & NO            &               & YES           &               & NO            &               & YES            \Tstrut\Bstrut\\
                          
        $R^2$             &               & $0.277$       &               & $0.525$       &               & $0.228$       &               & $0.477$       \Tstrut\Bstrut \\
      
        $N$               &               & $27,390$      &               & $27,390$      &               & $19,384$      &               & $19,384$ 
         \Tstrut\Bstrut\\
        \bottomrule
        \end{tabular}
        \begin{tablenotes}[flushleft]
        \linespread{1}\footnotesize
        \item\hspace*{-\fontdimen2\font}\textit{Notes:} Standard errors clustered at the industry level are in parenthesis. 
        $^{***}$, $^{**}$ and $^*$ indicate significance at the $1\%$, $5\%$ and $10\%$ levels.
        \end{tablenotes}
    \end{threeparttable}    
\end{table}

Table~(\ref{tbl:heterResp}) allows to compute the average impact of a shock on the current RDI of different firms. For example, based on the first column, firms in the third quartile ($ECDF=0.75$) respond to the shock by increasing their RDI by $-0.188+0.639\times 0.75=29\%$. We repeat the procedures for the other three columns and firms in the first quartile ($ECDF=0.25$). Table~(\ref{tbl:heterResp7525}) reports the results: across all specifications, low--cost firms increase RDI and high--cost do the opposite. The results are quantitatively similar to our model predictions: Figure~(\ref{fig:sim}) shows that the low-cost firm increases R\&D effort by $23\%$ while the high-cost firm exhibits an instant drop in by $7\%$. 
\begin{table}[!htbp]
    \centering
    \begin{threeparttable}
        \caption{Third and First Quartile Firm Responses} \label{tbl:heterResp7525}
        \begin{tabular}{lcccccccc}
        \hline\hline
        \quad  & (1)    & \qquad & (2)    & \qquad & (3)    & \qquad & (4)            \Tstrut\Bstrut  \\
        \hline
        First quartile firms  & $-0.028$      & & $-0.082^{**}$ & & $-0.103$      & & $-0.189^{**}$   \Tstrut\Bstrut \\
               & (0.037)       & & (0.037)       & & (0.066)       & & (0.086)       \Tstrut\Bstrut   \\
        Third quartile firms  & $0.291^{***}$ & & $0.179^{***}$ & & $0.293^{***}$ & & $0.204^{***}$   \Tstrut\Bstrut \\
               & (0.072)       & & (0.037)       & & (0.101)       & & (0.069)       \Tstrut\Bstrut   \\
               &               & &               & &               & &               \Tstrut\Bstrut   \\
ECDF measure   & L1            & & L1            & & MA3           & & MA3            \Tstrut\Bstrut  \\              
Control variables & NO         & & YES           & & NO            & & YES             \Tstrut\Bstrut \\              
        \bottomrule
        \end{tabular}
        \begin{tablenotes}[flushleft]
        \linespread{1}\footnotesize
        \item\hspace*{-\fontdimen2\font}\textit{Notes:} Standard errors clustered at the industry level are in parenthesis. 
        $^{***}$, $^{**}$ and $^*$ indicate significance at the $1\%$, $5\%$ and $10\%$ levels. L1 means we used $ECDF_{t-1}$ in the specification, and $MA3$ means we used $ECDF^{MA3}_{t-1}$ instead. 
        \end{tablenotes}
    \end{threeparttable}    
\end{table}

Alternatively, we construct the shocks using 1\% and 10\% as thresholds and compute how firms respond to different magnitudes of shocks.  Table~(\ref{tbl:heterResp7525_robust}) in Appendix~\ref{app: robustness} tabulate the results. The takeaway is that the more drastic the shock, the more heterogeneous the responses: the first--quartile firms reduce more RDI, while the third--quartile firms do the opposite. 
%%%%%%%%%%%%%%%%%%%%%%%%%%%%%%%%%%%%%%%%%%%%%%%%%%
\subsection{Sorting Approach}
\begin{figure}[ht]
  \centering
  \caption{Groupwise responses of firm RDI to the NPS}\label{fig:groupReg}
  \includegraphics[width=1\textwidth]{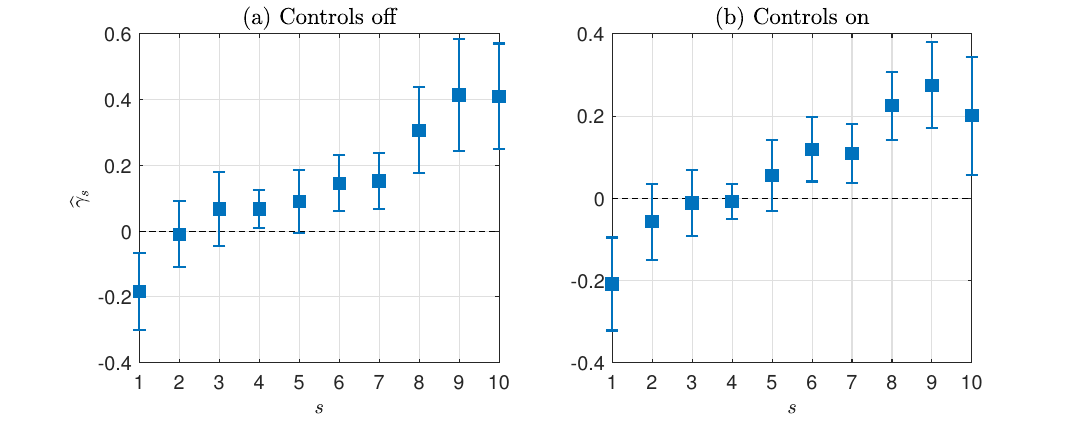}
  \begin{minipage}{0.8\textwidth}
	{\footnotesize Note: Panel (a) plots the groupwise responses of firm-level RDI to the negative profit shock ($\left\{\widehat{\gamma}^s\right\}_{s=1}^{10}$) and the $90\%$ confidence intervals, using no control variables; panel (b) plots those using control variables. The CIs are based on robust standard errors clustered on the industry level. \par}
  \end{minipage}
\end{figure}

The coefficient $\gamma$ in specification~(\ref{eqn:regSpec1}) is linear in past RDIs, so we consider an alternative specification that allows for non--linear R\&D responses:
\begin{equation}\label{eqn:regSpec2}
    \log(RDI_{ijt}) = \alpha_i + \delta_t + \sum_{s=1}^{10}\gamma_s NPS_{j,t-1}\times c_{i,j,t-1}^s + \sum_{s=1}^{10}\eta_s c_{i,j,t-1}^s + \boldsymbol{\lambda}\boldsymbol{X}_{ijt} + u_{ijt}.
\end{equation}
We do not include $NPS_{j,t-1}$ because the group of cross-dummies $\left\{NPS_{jt}\times c_{ijt}^s\right\}_{s=1}^{10}$ fully saturates the specification already. Figure~(\ref{fig:groupReg}) plots $\{\widehat{\gamma}_{s}\}^{10}_{s=1}$, estimators of the group--wise responses of RDI to the shocks, and $90\%$ confidence intervals. The general pattern is consistent with the model prediction, namely figure~(\ref{fig:sim}), that RDI responses are increasing in group deciles. In particular, firms in the first two deciles decrease RDI in response to the shock.We also consider shocks with 1\% and 10\% as thresholds. Figure~(\ref{fig:groupReg_pc1}) and (\ref{fig:groupReg_pc10}) plot the results using the sorting approach. The general patterns of heterogeneous responses still hold, and the responses to a more drastic shock (1\%) is more spread out. Overall, the two robustness checks provide supporting evidence that firms respond differently to different magnitudes of 

%\indent The interpretation of $\widehat{\gamma}_s$ in specification (\ref{eqn:regSpec2}) is similar to $\widehat{\beta}+\widehat{\gamma}$ in (\ref{eqn:regSpec1}). It is the average impact of the negative profit shock on RDI of firms in the decile group $s$. 

To summarize, we employ two regression approaches to test the model prediction that negative profit shocks induce heterogeneous R\&D responses. We find that firms with high RDI in the past respond to the shock by raising their RDI, while those with low past RDI do the opposite. The two approaches produce comparable sizes of R\&D responses, which are also close to results obtained from the model simulations. These findings suggest that our mechanism can explain a sizable fraction of the countercyclical productivity dispersion. 
%%%%%%%%%%%%%%%%%%%%%%%%%%%%%%%%%%%%%%%%%%%%%%%%
%%%%%%%%%%%%%%%%%%%%%%%%%%%%%%%%%%%%%%%%%%%%%%%%
\section{Conclusion}\label{sec:conc}
This paper proposes a new theory for the cause of countercyclical productivity dispersion. We create a measure of industry--level negative profit shocks and document their dynamic enlargement effects on productivity dispersion and R\&D intensity dispersion. Through the lens of a duopolistic technology ladder model, we establish a mechanism in which firms' heterogeneous R\&D responses to the shock leads to a wider technology gap in equilibrium, hence a larger productivity dispersion. In the quantitative analysis, we fit the paramaterized model to data and predict that the firm with the lower R\&D cost increases its research effort on average after a sizable shock to profit, while the firm with the higher cost does the opposite. The simulation accounts for $60\%$ of the observed productivity dispersion in response to the shock. We provide empirical evidence to support the existence of this mechanism in the data. 

We see three important avenues for future research. First, throughout the paper we are agnostic about the source of the negative profit shocks and assume that such shocks affect firms equally. Providing a micro foundation of this shock to profit (e.g., a technology shock, a raw material price shock, etc.) could shed light on potential heterogeneous impacts on firms of different characteristics. 

Second, a recent paper by \cite{BV2019} uses exchange rate movements as an observable cost shock to disentangle the contributions of changes in firms' responses and changes in shocks' volatility to the greater dispersion of an endogenous variable, i.e., item--level price. In this paper, we show that firms' heterogeneous responses to negative profit shocks causes a greater dispersion in the productivity gap, which is endogenous, but we are silent on the volatility changes of the shocks. It is worth investigating whether these negative profit shocks experience changes in higher--order moments over time and what their impacts on the productivity dispersion might be. 

Third, the model has several simplifying assumptions. We assume there is a maximum technology gap, but it can potentially be endogenized if the model allows for firm entry and exit: an incumbent exits if it is left too far behind, and a new firm can decide whether it wants to enter. Due to a lack of entry--exit data, we do not study how negative profit shocks affect entry--exit dynamics and the implications for productivity dispersion, but these are interesting questions both empirically and theoretically. On the other hand, the model assumes that the leader and the laggard innovate with the same quality, but there is ample evidence that innovation quality differs across firms \citep{AkcigitKerr2018, Olmstead-Rumsey2020}. Allowing for heterogeneous R\&D quality might affect firms' effort decisions and potentially amplify the channel. 

%Finally, the new mechanism in this paper serves to explain countercyclical productivity dispersion. It would be interesting to build a theoretical framework, in the spirit of \cite{AkcigitAtes2019} and \cite{CavenaileCelikTian2021}, that nests different mechanisms mentioned in the literature review and quantifies the explanatory power of each channel. 

%How do we link to growth of economy? \textcolor{winered}{[Ming: In our structurally estimated model, in response to a NPS, the high-cost firm reduces its R\&D effort, but exhibit positive productivity growth. This is because in the stationary equilibrium, for the most of the time the leader is at the maximum gap. Therefore, when there is a shock, the spillover at the maximum gap effect dominates the reduction in R\&D, and leads to positive IRF of productivity growth of both two firms. This is why I think it may attract doubts and critiques.]} I think this is a quantitative issue since $\overline{m}$ is set to be 2. So it is very likely that the leader is going to be at the max gap. On the other hand, if $\overline{m}$ is large, then the model doesn't generate quantitatively large response, which cannot explain the IRF from the local projection. \textcolor{red}{[SC: We need to strike a balance by setting an $\overline{m}$ reasonable.]}

\newpage
\bibliographystyle{aea}
\bibliography{JMP_Ming} 

\newpage
\section*{Appendices}
\begin{appendices}
\section{Data Appendix}\label{asc:data}
\setcounter{equation}{0}
\numberwithin{equation}{section}
\renewcommand{\theequation}{A.\arabic{equation}}
%%%%%%%%%%%%%%%%%%%%%%%%%%%%%%%%%%%%%%%%%%%%%%%%
\subsection{Sample Selection} 
For the CRSP/Compustat merged data, we restrict the sample to U.S–based firms that provide final versions of statements. We omit regulated utilities (SIC codes 4900 to 5000) and financial firms (SIC codes 6000 to 7000), get rid of firm-year observations with values of acquisitions greater than 5\% of assets, and keep only if the firm exists in the data for at least two years. We also drop observations with negative or missing book value of assets, book value of capital, number of employees, capital investment or revenue. Because Compustat records end-of-year captal values, we shift the reported book value forward one year.

For each industry definded by a four-digit SIC code and year in the NBER-CES database, we compute the following two variables: the labor share in value added (payroll cost divided by value added, with variable name \texttt{labshare}), the ratio of value added to gross output (\texttt{vaddfrac}). We then replace these two variables by their respective 10-year moving average, and generate the capital share (\texttt{capshare}) as the residual of the labor share, where we make the underlying assumption that the production function is homogeneous of degree one in labor and capital.

We merge the CRSP/Compustat and NBER-CES Datasets by industry and year indicator (\texttt{gvkey} and \texttt{year}, respectively), and then merge with it the \cite{KPSS2017} firm innovation value dataset by the permanent company and security identifier (\texttt{permno}) and \texttt{year}. This yields an unbalanced panel dataset, whose time spans annualy from 1970 to 2010, and covers 4,074 firms (identified by Compustat’s permanent company identifier, \texttt{gvkey}) out of 135 four-digit SIC industries. There are 43,800 observations in total.
%%%%%%%%%%%%%%%%%%%%%%%%%%%%%%%%%%%%%%%%%%%%%%%%
\subsection{Construction of Key Variables}\label{sec:keyVar}
\subsubsection*{Revenue Total Factor Productivity (TFPR)}

The revenue TFP in this paper is estimated using the cost-share based approach by \cite{FHS2008}.\footnote{\cite{HK2009} use plant--level marginal revenue of product capital and labor (MPRK and MPRL) to construct total factor productivity, while \cite{KV2020} use MRPK only. See \cite{Haltiwanger2016} for a discussion of the relative merits of the approaches.} We first define variable \texttt{cap} as the book value of capital (\texttt{ppent} in CCM) deflated by the investment deflator (\texttt{piinv} from the NBER-CES database). Then we define variable \texttt{output} as \texttt{sale} multiplied by the value added to gross output ratio (\texttt{vaddfrac}), which is then deflated by the shipments deflator (\texttt{piship}). The log TFPR is calculated as follows:
\begin{equation}\label{app:tfpr}
    \log{TFP}_{ijt} = \log(output_{ijt})-capshare_{jt}\log(cap_{ijt})-labshare_{jt}\log(emp_{ijt}).
\end{equation}

As such, \texttt{log\_tfp} is the residual of revenue that is not explained by the factors capital and labor in a production function homogeneous of degree one, whose factor shares are invariant across firms in each industry.

\subsubsection*{R\&D Expenditure (RDE) and Intensity (RDI)}

The CCM database provides firm-year observations of R\&D expenses (\texttt{xrd}), and we scale \texttt{xrd} by firm size, approximated by sale, to get the RDI. 

In Section \ref{sec:emp}, we use the natural log of RDI as the explained variable instead of the level. This is because the latter is highly right-skewed, with mean $4.49$ and the maximum as high as $25684.40$. This may be due to the fact some firms may have sales close to zero at times when their RDE is far from zero. 
%%%%%%%%%%%%%%%%%%%%%%%%%%%%%%%%%%%%
% The subsection below is added by Yang Ming
\subsection{Estimating the Impact of NPS}\label{appendix:estNPS}
We run two reduced-form regressions to estimate the impacts of negative profit shocks on (1) firms' aggregate profits and (2) the interquartile range of TFP at the industry-year level. To compare them with those in our simulated model.
\par
Let $Y_{j,t}$ be the explained variable, either the natural log of gross profit aggregated at the industry-year level ($\log{gp}_{j,t}$), or the natural log of interquartile range of TFP at the same level ($\log{TFP\_iqr}_{j,t}$). The regression specification is as follows:
\begin{equation}\label{app:regSpec}
    Y_{j,t} = \alpha_j + \delta_t + \beta NPS_{j,t} + \gamma_1 nFirm_{j,t-1} + \gamma_2 HHI_{j,t-1} + \varepsilon_{j,t},
\end{equation}
where $\alpha_j$ and $\delta_t$ are the industry and year fixed effects, respectively; $NPS$ is the negative profit shock defined in (\ref{eqn:npsDef}). We include one-period lag of industry-year firm number ($nFirm$) and Herfindahl-Hirschman Index ($HHI$) as covariates to control the sector-time variation in market structure. The results are reported in the table below:
\begin{table}[htbp!]
\centering
\begin{threeparttable}
  \caption{Contemporaneous Effects of the NPS}\label{tbl:effNPS}
  \begin{tabular}{lccccc}
  \hline\hline
                    & \qquad & $\log{gp}$        & \qquad & $\log{TFP\_iqr}$     \Tstrut\Bstrut  \\
  \hline
  $NPS$             & \qquad & $-0.192^{**}$     &        & $0.096^{***}$         \Tstrut\Bstrut \\
                    & \qquad & $(0.082)$         &        & $(0.035)$           \Tstrut\Bstrut   \\
                    & \qquad &                   &        &                 \Tstrut\Bstrut       \\
  $nFirm$           & \qquad & $0.043^{***}$     &        & $0.006^{***}$        \Tstrut\Bstrut  \\ 
                    & \qquad & $(0.010)$         &        & $(0.001)$       \Tstrut\Bstrut       \\  
  $HHI$             & \qquad & $-0.091$          &        & $0.169^*$          \Tstrut\Bstrut    \\
                    & \qquad & $(0.245)$         &        & $(0.094)$         \Tstrut\Bstrut     \\  
                    & \qquad &                   &        &                   \Tstrut\Bstrut     \\
  $R^2$             & \qquad & $0.287$           &        & $0.126$          \Tstrut\Bstrut      \\
  $N$               & \qquad & $4,164$           &        & $4.172$            \Tstrut\Bstrut    \\
  \bottomrule
  \end{tabular}
  \begin{tablenotes}[flushleft]
  \linespread{1}\footnotesize
  \item\hspace*{-\fontdimen2\font}\textit{Notes:} The negative profit shock (NPS) is a dummy variable which obtains value $1$ if the industry-year profit falls below its 10th percentile. Observations with less than $2$ firms in the industry-year cell are excluded. Standard errors clustered at the industry level are in parenthesis. 
  $^{***}$, $^{**}$ and $^*$ indicate significance at the $1\%$, $5\%$ and $10\%$ levels.
  \end{tablenotes}
\end{threeparttable}      
\end{table}

%%%%%%%%%%%%%%%%%%%%%%%%%%%%%%%%%%%%
\subsection{Results of NPS with Different Thresholds} \label{app: robustness}
\begin{figure}[ht]
  \centering
  \caption{Groupwise responses of firm RDI to NPS at 1\% Threshold}\label{fig:groupReg_pc1}
  \includegraphics[width=1\textwidth]{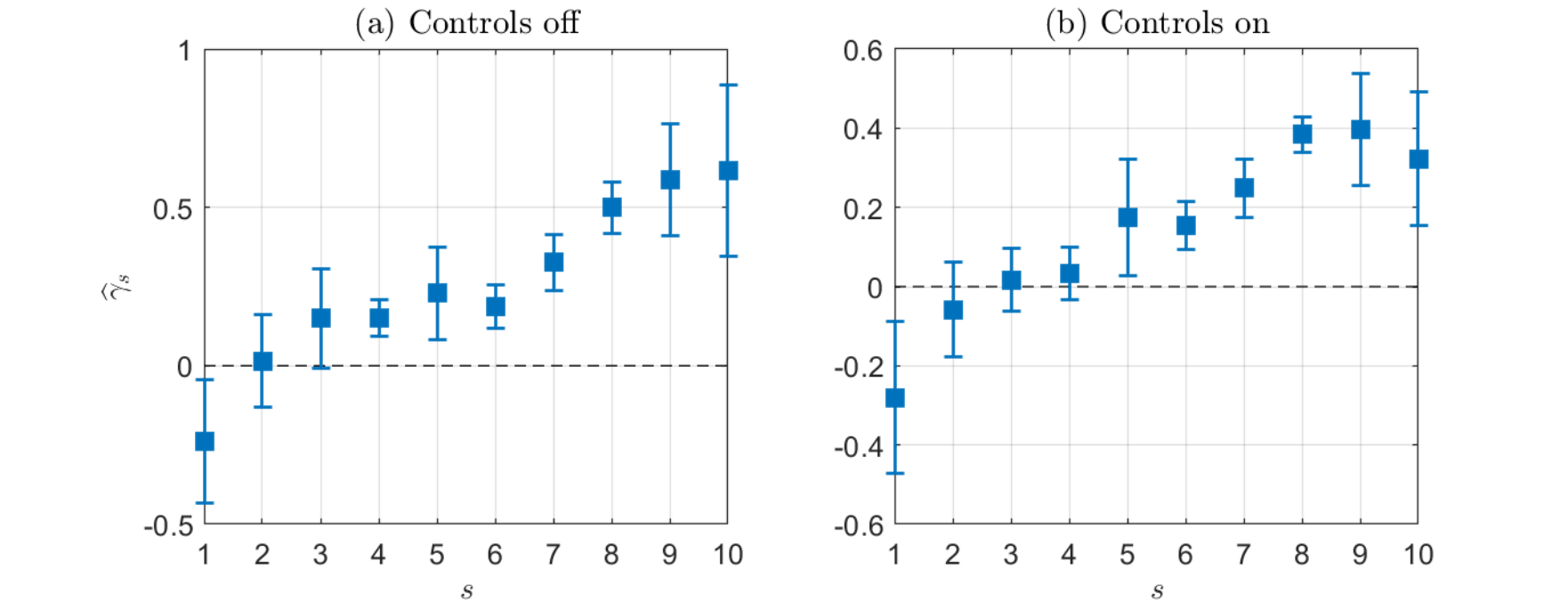}
  \begin{minipage}{0.8\textwidth}
	{\footnotesize Note: Panel (a) plots the groupwise responses of firm-level RDI to the negative profit shock ($\left\{\widehat{\gamma}^s\right\}_{s=1}^{10}$) and the $90\%$ confidence intervals, using no control variables; panel (b) plots those using control variables. The CIs are based on robust standard errors clustered on the industry level. \par}
  \end{minipage}
\end{figure}

\begin{figure}[ht]
  \centering
  \caption{Groupwise responses of firm RDI to NPS at 10\% Threshold}\label{fig:groupReg_pc10}
  \includegraphics[width=1\textwidth]{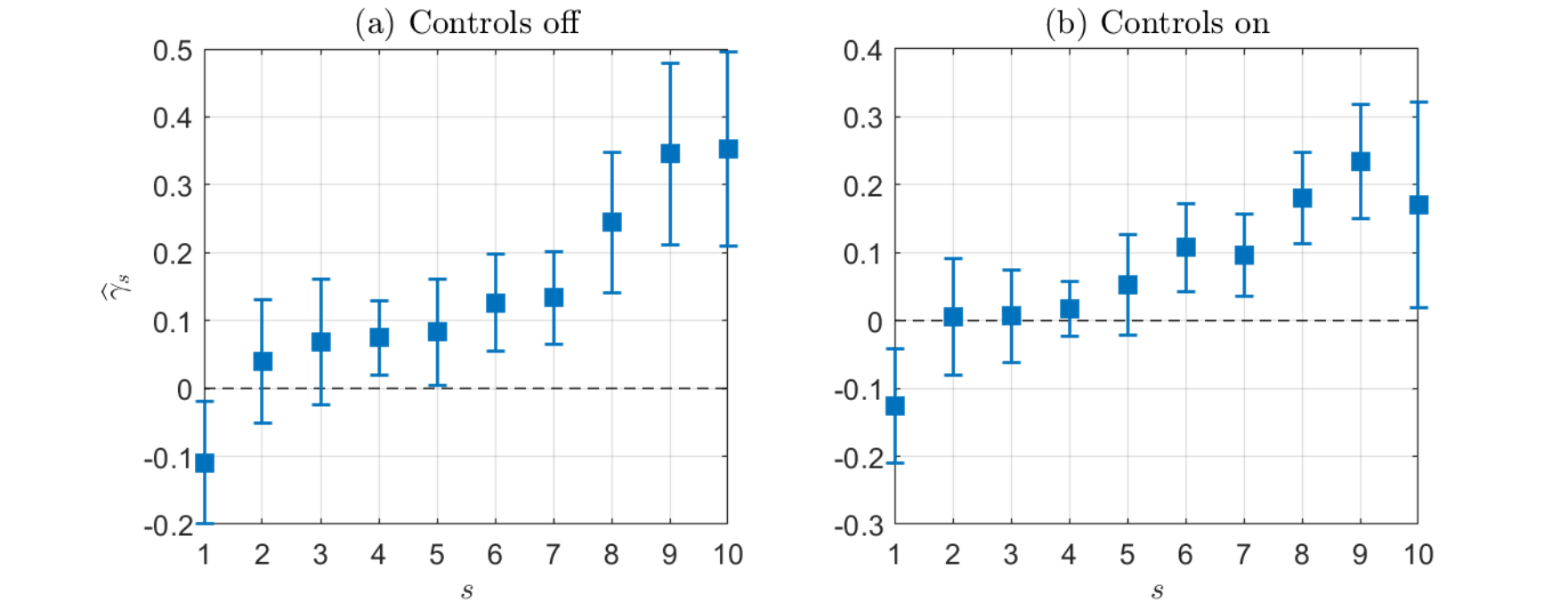}
  \begin{minipage}{0.8\textwidth}
	{\footnotesize Note: Panel (a) plots the groupwise responses of firm-level RDI to the negative profit shock ($\left\{\widehat{\gamma}^s\right\}_{s=1}^{10}$) and the $90\%$ confidence intervals, using no control variables; panel (b) plots those using control variables. The CIs are based on robust standard errors clustered on the industry level. \par}
  \end{minipage}
\end{figure}

The main text constructs the negative profit shocks (NPS) using 5\% as the threshold. Alternatively, we consider 1\% and 10\% as thresholds and redo the empirical analysis in Section~\ref{sec:emp}. Tables~(\ref{tbl:robustCheck_pct1}) and (\ref{tbl:robustCheck_pct10}) show the results of ECDF approach. Compared to Table~(\ref{tbl:heterResp7525}), firms' responses are more different to a more drastic shock (1\%). Figure~(\ref{fig:groupReg_pc1}) and (\ref{fig:groupReg_pc10}) plot the results using the sorting approach. The general patterns of heterogeneous responses still hold, and the responses to a more drastic shock (1\%) is more spread out. Overall, the two robustness checks provide supporting evidence that firms respond differently to different magnitudes of NPS.
\begin{table}[htbp!]
\centering
\begin{threeparttable}
\centering
  \caption{Responses of R\&D intensity to NPS at 1\% Threshold} \label{tbl:robustCheck_pct1}
  \begin{tabular}{lcccccccc}
  \hline\hline
  $\log(RDI_t)$     & \quad         & (1)           & \qquad        & (2)           & \qquad        & (3)           & \qquad        & (4)        \Tstrut\Bstrut    \\
  \hline
  $NPS_{t-1}$       &               & $-0.201^{**}$ &               & $-0.247^{***}$&               & $-0.426^{***}$&               & $-0.557^{***}$  \Tstrut\Bstrut\\
                    &               & $(0.098)$     &               & $(0.080)$     &               & $(0.153)$     &               & $(0.164)$  \Tstrut\Bstrut    \\
  $NPS_{t-1}\times ECDF_{t-1}$
                    &               & $0.909^{***}$ &               & $0.739^{***}$ &               &               &               &        \Tstrut\Bstrut  \\
                    &               & $(0.223)$     &               & $(0.143)$     &               &               &               &           \Tstrut\Bstrut     \\
  $ECDF_{t-1}$
                    &               & $1.400^{***}$ &               & $1.292^{***}$ &               &               &               &             \Tstrut\Bstrut   \\
                    &               & $(0.146)$     &               & $(0.104)$     &               &               &               &             \Tstrut\Bstrut   \\
  $NPS_{t-1}\times ECDF_{t-1}^\text{MA3}$
                    &               &               &               &               &               & $1.391^{***}$ &               & $1.376^{***}$ \Tstrut\Bstrut \\
                    &               &               &               &               &               & $(0.361)$     &               & $(0.333)$    \Tstrut\Bstrut  \\
  $ECDF_{t-1}^\text{MA3}$
                    &               &               &               &               &               & $1.149^{***}$ &               & $1.146^{***}$  \Tstrut\Bstrut\\
                    &               &               &               &               &               & $(0.184)$     &               & $(0.176)$   \Tstrut\Bstrut   \\                                        
                    &               &               &               &               &               &               &               &              \Tstrut\Bstrut  \\
  Control variables &               & NO            &               & YES           &               & NO            &               & YES         \Tstrut\Bstrut   \\
                    
  $R^2$             &               & $0.281$       &               & $0.521$       &               & $0.251$       &               & $0.475$     \Tstrut\Bstrut   \\

  $N$               &               & $27,390$      &               & $27,390$      &               & $19,384$      &               & $19,384$     \Tstrut\Bstrut  \\
  \bottomrule
  \end{tabular}
  \begin{tablenotes}[flushleft]
  \linespread{1}\footnotesize
  \item\hspace*{-\fontdimen2\font}\textit{Notes:} The negative profit shock (NPS) is a dummy variable which obtains value $1$ if the industry-year profit falls below its 1st percentile. Standard errors clustered at the industry level are in parenthesis. 
  $^{***}$, $^{**}$ and $^*$ indicate significance at the $1\%$, $5\%$ and $10\%$ levels.
  \end{tablenotes}
\end{threeparttable}
\end{table}
\clearpage
\begin{table}[htbp!]
\centering
\begin{threeparttable}
  \caption{Responses of R\&D intensity to NPS at 10\% Threshold}\label{tbl:robustCheck_pct10}
  \begin{tabular}{lcccccccc}
  \hline\hline
  $\log(RDI_t)$     & \quad         & (1)           & \qquad        & (2)           & \qquad        & (3)           & \qquad        & (4)          \Tstrut\Bstrut  \\
  \hline
  $NPS_{t-1}$       &               & $-0.113^{*}$  &               & $-0.126^{**}$ &               & $-0.193^{*}$  &               & $-0.259^{**}$  \Tstrut\Bstrut\\
                    &               & $(0.064)$     &               & $(0.053)$     &               & $(0.109)$     &               & $(0.123)$     \Tstrut\Bstrut \\
  $NPS_{t-1}\times ECDF_{t-1}$
                    &               & $0.466^{***}$ &               & $0.364^{***}$ &               &               &               &             \Tstrut\Bstrut   \\
                    &               & $(0.137)$     &               & $(0.087)$     &               &               &               &             \Tstrut\Bstrut   \\
  $ECDF_{t-1}$
                    &               & $1.343^{***}$ &               & $1.246^{***}$ &               &               &               &              \Tstrut\Bstrut  \\
                    &               & $(0.134)$     &               & $(0.099)$     &               &               &               &              \Tstrut\Bstrut  \\
  $NPS_{t-1}\times ECDF_{t-1}^\text{MA3}$
                    &               &               &               &               &               & $0.561^{**}$  &               & $0.560^{**}$  \Tstrut\Bstrut \\
                    &               &               &               &               &               & $(0.225)$     &               & $(0.216)$     \Tstrut\Bstrut \\
  $ECDF_{t-1}^\text{MA3}$
                    &               &               &               &               &               & $1.092^{***}$ &               & $1.087^{***}$  \Tstrut\Bstrut\\
                    &               &               &               &               &               & $(0.162)$     &               & $(0.155)$     \Tstrut\Bstrut \\                                        
                    &               &               &               &               &               &               &               &              \Tstrut\Bstrut  \\
  Control variables &               & NO            &               & YES           &               & NO            &               & YES          \Tstrut\Bstrut  \\
                    
  $R^2$             &               & $0.265$       &               & $0.515$       &               & $0.212$       &               & $0.467$      \Tstrut\Bstrut  \\

  $N$               &               & $27,390$      &               & $27,390$      &               & $19,384$      &               & $19,384$    \Tstrut\Bstrut   \\
  \bottomrule
  \end{tabular}
  \begin{tablenotes}[flushleft]
  \linespread{1}\footnotesize
  \item\hspace*{-\fontdimen2\font}\textit{Notes:} The negative profit shock (NPS) is a dummy variable which obtains value $1$ if the industry-year profit falls below its 10th percentile. Standard errors clustered at the industry level are in parenthesis. 
  $^{***}$, $^{**}$ and $^*$ indicate significance at the $1\%$, $5\%$ and $10\%$ levels.
  \end{tablenotes}
\end{threeparttable}      
\end{table}

\begin{table}[!htbp]
    \centering
    \begin{threeparttable}
        \caption{Third and First Quartile Firm Responses to NPS with Alternative Thresholds} \label{tbl:heterResp7525_robust}
        \begin{tabular}{lcccccccc}
        \hline\hline
               & (1)    & \qquad & (2)    & \qquad & (3)    & \qquad & (4)         \Tstrut\Bstrut     \\
        \hline
        NPS at $1\%$ Threshold                                                     \Tstrut\Bstrut     \\
        First quartile firms  & $0.026$       & & $-0.062$      & & $-0.078$      & & $-0.214^{**}$   \Tstrut\Bstrut \\
               & $(0.053)$     & & $(0.048)$     & & $(0.072)$     & & $(0.084)$    \Tstrut\Bstrut    \\
        Third quartile firms  & $0.480^{***}$ & & $0.307^{***}$ & & $0.617^{***}$ & & $0.474^{***}$   \Tstrut\Bstrut \\
               & $(0.088)$     & & $(0.041)$     & & $(0.132)$     & & $(0.092)$    \Tstrut\Bstrut    \\
               &               & &               & &               & &        \Tstrut\Bstrut          \\
        NPS at $10\%$ Threshold                                                     \Tstrut\Bstrut     \\
        First quartile firms  & $0.003$       & & $-0.035$      & & $-0.053$      & & $-0.119^*$   \Tstrut\Bstrut    \\
               & $(0.036)$     & & $(0.035)$     & & $(0.057)$     & & $(0.071)$   \Tstrut\Bstrut     \\
        Third quartile firms  & $0.236^{***}$ & & $0.147$       & & $0.228^{***}$ & & $0.161^{***}$  \Tstrut\Bstrut  \\
               & $(0.053)$     & & $(0.030)$     & & $(0.071)$     & & $(0.048)$    \Tstrut\Bstrut    \\
               &               & &               & &               & &     \Tstrut\Bstrut             \\
ECDF measure   & L1            & & L1            & & MA3           & & MA3          \Tstrut\Bstrut    \\              
Control variables & NO         & & YES           & & NO            & & YES           \Tstrut\Bstrut   \\
        \bottomrule
        \end{tabular}
        \begin{tablenotes}[flushleft]
        \linespread{1}\footnotesize
        \item\hspace*{-\fontdimen2\font}\textit{Notes:} Standard errors clustered at the industry level are in parenthesis. 
        $^{***}$, $^{**}$ and $^*$ indicate significance at the $1\%$, $5\%$ and $10\%$ levels. L1 means we used $ECDF_{t-1}$ in the specification, and $MA3$ means we used $ECDF^{MA3}_{t-1}$ instead. 
        \end{tablenotes}
    \end{threeparttable}    
\end{table}

\FloatBarrier

%%%%%%%%%%%%%%%%%%%%%%%%%%%%%%%%%%%%
%%%%%%%%%%%%%%%%%%%%%%%%%%%%%%%%%%%%
\clearpage
\newpage
\section{Proof Appendix} \label{Proof Appendix}
\setcounter{equation}{0}
\numberwithin{equation}{section}
\renewcommand{\theequation}{B.\arabic{equation}}
%%%%%%%%%%%%%%%%%%%%%%%%%%%%%%%%%%%%
\subsection{Proof of Lemma \ref{lem:mkvStr}}
\begin{proof}
Fix $s\geq 0$, and let $t\neq s$ be such that $\Delta n_f(s)=\Delta n_f(t)$. Since firm $-f$ plays a $\Delta n_{-f}$-dependent strategy, by Assumption \ref{asm:pi} and performance measure (\ref{eqn:pm}), firm $f$ faces the same optimization problem. Thus by the strict convexity of R\&D cost function, it implies that the set $\left\{t \mid s\neq t, \Delta n_f(s)=\Delta n_f(t), a_f^*(s)\neq a_f^*(t)\right\}$ has zero measure, and it must be empty set if $a_f^*(t)$ is right-continuous.
\end{proof}
%%%%%%%%%%%%%%%%%%%%%%%%%%%%%%%%%%%%
\subsection{Proof of Proposition \ref{prp:exist}}\label{prf:prp:exist}
\begin{proof}
First simplify Equation (\ref{eqn:valFcnMpe}) by removing the expectation operator. Note that the composite arriving time $Z\in(0,\infty)$ follows an exponential distribution:
\begin{align}
    F_Z(z) =& 1 - \pr{(Z>z)} \nonumber  \\
           =& 1 - \pr{(Z_A>z,Z_B>z)} \nonumber \\
           = &1 - \exp{\left[-\left(\lambda(a_A^*+a_B^*)+h\cdot\mathds{1}\{\Delta n_i \neq 0\}\right)z\right]}. \label{app:zCdf}
\end{align}
Let $Y = e^{-\rho Z}<1$, then it can be shown that $Y$ follows a Beta distribution: 
$$Y\sim \text{Beta}(\ol{\lambda}/\rho,1),$$ 
where $\ol{\lambda}\coloneqq \lambda(a_A^*+a_B^*)+h\cdot\mathds{1}\{\Delta n_i \neq 0\}$. The CDF and PDF are
\begin{align}
    F_Y(y) =& \pr{\left(Z\leq -\dfrac{\ln{y}}{\rho}\right)} = \int_0^{\scalebox{0.75}{$-\dfrac{\ln{y}}{\rho}$}} \ol{\lambda}\exp(-\ol{\lambda}z) dz = 1 - y^{\ol{\lambda}/\rho}; \label{app:yCdf} \\
    f_Y(y) =& -\dfrac{\ol{\lambda}}{\rho}y^{(\ol{\lambda}/\rho)-1}
    = \dfrac{1}{\text{B}(\ol{\lambda}/\rho,1)}y^{(\ol{\lambda}/\rho)-1}, \label{app:yPdf}
\end{align}
where $\text{B}(x,y)=\int_0^1 t^{x-1}(1-t)^{y-1}dt$ is the Beta function. Plugging the expressions into Equation~(\ref{eqn:valFcnMpe}), we have
\begin{align}
    \mathds{E}_Z\left[\int_0^Z \exp(-\rho t)dt\right]
    =& \dfrac{1}{\rho}\left[1-\mathds{E}_Z\left(\exp(-\rho Z)\right)\right]\nonumber \\
    =& \dfrac{1}{\rho}\left[1-\mathds{E}_Y(Y)\right] \nonumber \\
    =& \dfrac{1}{\lambda(a_A^*+a_B^*)+h\cdot\mathds{1}\{\Delta n_i \neq 0\} + \rho}; \label{app:exInt} \\
    \mathds{E}_Z\left[\exp(-\rho Z)\mathds{1}\{Z=Z_i\}\right]
    =& \mathds{E}_Z\left[\exp(-\rho Z)\right]\mathds{E}_Z\left[\mathds{1}\{Z=Z_i\}\right] \nonumber \\
    =& \mathds{E}_Y(Y)\pr{(Z_i<Z_{-i})} \nonumber \\
    =& \dfrac{\lambda(a_A^*+a_B^*)+h\cdot\mathds{1}\{\Delta n_i \neq 0\}}{\lambda(a_A^*+a_B^*)+h\cdot\mathds{1}\{\Delta n_i \neq 0\} + \rho} \cdot
    \dfrac{\lambda a_i^* + h\cdot\mathds{1}\{\Delta n_i < 0\}}{\lambda(a_A^*+a_B^*)+h\cdot\mathds{1}\{\Delta n_i \neq 0\}} \nonumber \\
    =& \dfrac{\lambda a_i^* + h\cdot\mathds{1}\{\Delta n_i < 0\}}{\lambda(a_A^*+a_B^*)+h\cdot\mathds{1}\{\Delta n_i \neq 0\} + \rho}. \label{app:exExp}
\end{align}

Substituting Equation (\ref{app:exInt}) and (\ref{app:exExp}) into Bellman equation (\ref{eqn:valFcnMpe}), we derive the optimization system as in Proposition \ref{prp:exist}. The boundary conditions are due to Assumption~\ref{asm:automaticgap}, which implies that a firm has no incentive to do R\&D at the maximal technology gap $\ol{m}$.

The existence of a Markov Perfect Equilibrium is guaranteed Kakutani's Fixed-Point Theorem. We directly apply Theorem 5.11.15 in \cite{CSZ2009}.
\end{proof}
%%%%%%%%%%%%%%%%%%%%%%%%%%%%%%%%%%%%
\subsection{Proof of Proposition \ref{prp:statDist}}
\begin{proof}
Theorems 3.5.1 and 3.5.2 in \cite{NORRIS1998} establish the existence and uniqueness of the \textit{invariant measure}, which has a one--to--one mapping to the limiting distribution.
\end{proof}
%%%%%%%%%%%%%%%%%%%%%%%%%%%%%%%%%%%%
\subsection{Proof of Proposition \ref{prp:gapHomo}}
\begin{proof}
By the symmetry of the game, firms $A$ and $B$ have the same value function and policy function; otherwise it is easy to show there is contradiction by switching firm labels. As such, $\left\{\Delta n_A(t)\right\}$ and $\left\{\Delta n_B(t)\right\}$ have the same limiting distribution $\mu$. By definition $\Delta n_A(t)=-\Delta n_B(t)$. Therefore $\forall m\in\mathcal{M}$, 
$$\displaystyle\mu_i=\lim_{t\rightarrow\infty}\pr{\left(\Delta n_A(t)=m\right)}=\lim_{t\rightarrow\infty}\pr{\left(\Delta n_B(t)=-m\right)}=\lim_{t\rightarrow\infty}\pr{\left(\Delta n_A(t)=-m\right)}=\mu_{2\ol{m}+2-i}.$$ 
Since $\mathcal{M}_i=-\mathcal{M}_{2\ol{m}+2-i}$,  $$\displaystyle\lim_{t\rightarrow\infty}\mathds{E}\left[\Delta n_A(t)\right]=\sum_{i=1}^{2\ol{m}+1}\mu_i\mathcal{M}_i=0.$$
\end{proof}
%%%%%%%%%%%%%%%%%%%%%%%%%%%%%%%%%%%%
\subsection{Proof of Lemma \ref{lem:valFcn}}
\begin{proof}
We prove by contradiction. Suppose $\exists m_0\in\mathcal{M}$ such that $v_A(m_0)\leq v_B(m_0)$. Then $\exists m$ such that $a_A^*(m)\neq a_B^*(m)$; otherwise $v_A(m)>v_B(m)$ $\forall m$ as firm $A$ has a lower marginal cost of R\&D. 
\\
\indent By Definition \ref{def:mpe}, denote the strategy profile by $\left\{a_A^*(\Delta n_A),a_B^*(\Delta n_B)\right\}$. Then $v_A(m_0)\leq v_B(m_0)$ implies that
\begin{align}
    &v_A\left(m_0\big|\left\{a_A^*,a_B^*\right\}\right) \nonumber \\
    =&\mathds{E}_{\Delta N_A}\left\{\int_0^\infty e^{-\rho t}\left[\pi_A\left(\Delta N_A(t)\right) - \psi_A\left(a_A^*\left(\Delta N_A(t)\right)\right)\right]dt\big|\Delta N_A(0)=m,\left\{a_A^*,a_B^*\right\}\right\} \nonumber \\
    \leq &\mathds{E}_{\Delta N_B}\left\{\int_0^\infty e^{-\rho t}\left[\pi_B\left(\Delta N_B(t)\right) - \psi_B\left(a_B^*\left(\Delta N_B(t)\right)\right)\right]dt\big|\Delta N_B(0)=m,\left\{a_A^*,a_B^*\right\}\right\} \nonumber \\
    =&v_B\left(m_0\big|\left\{a_A^*,a_B^*\right\}\right). \label{app:valFcnIneq}
\end{align}
Now let firm $A$ play $B$'s strategy, and denote firm $B$'s best response by $a_B^\text{br}$. We compare firm $A$'s original strategy $a_A^*$ and $a_B^\text{br}$, and there are three cases to consider:
\\
\textbf{Case 1:} $a_B^\text{br}(m)=a_A^*(m)$ for all $m$. In this case, it is easy to see from inequality (\ref{app:valFcnIneq}) that $$v_A\left(m_0 \mid \left\{a_B^*,a_B^\text{br}\right\}\right)>v_A\left(m_0 \mid \left\{a_A^*,a_B^*\right\}\right)$$ 
because the positions of the two firms are mirrored, and firm $A$ has strictly lower marginal cost than firm $B$.
\\
\textbf{Case 2:} $a_B^\text{br}(m)>a_A^*(m)$ for some $m$. This contradicts that equilibrium strategy $a_A^*$ is the best response to $a_B^*$. This is because firm $B$ faces the same problem when $A$ plays $a_B^*$ as $A$ faces in the original MPE. However, the marginal cost of R\&D is strictly higher for firm $B$, thus $a_B^\text{br}(m)>a_A^*(m)$ is impossible.
\\
\textbf{Case 3:} $a_B^\text{br}(m)\leq a_A^*(m)$ for all $m$, and this inequality holds strictly for some $m$. This implies that for any $t>0$ and any $k\in\{1,2,\cdots,\ol{m}\}$, 
\begin{align}
    \pr{\left(\Delta N_A(t)=k|\Delta N_A(0)=m;\left\{a_B^*,a_B^\text{br}\right\}\right)} 
    \geq&
    \pr{\left(\Delta N_A(t)=k|\Delta N_A(0)=m;\left\{a_B^*,a_A^*\right\}\right)} \nonumber \\
    =
    &\pr{\left(\Delta N_B(t)=k|\Delta N_B(0)=m;\left\{a_A^*,a_B^*\right\}\right)}, \label{app:probIneq1}
\end{align}
where the last equality is due to the symmetry of this dynamical system: once the initial states and R\&D are flipped and strategies of players $A$ and $B$ swapped, the random variable $\Delta N_f(t)$ is governed by the same stochastic process which $\Delta N_{-f}(t)$ initially follows. In a similar vein, 
\begin{equation}
    \pr{\left(\Delta N_A(t)=k|\Delta N_A(0)=m;\left\{a_B^*,a_B^\text{br}\right\}\right)} 
    \leq
    \pr{\left(\Delta N_B(t)=k|\Delta N_B(0)=m;\left\{a_A^*,a_B^*\right\}\right)} \label{app:probIneq2}
\end{equation}
for any $t>0$ and any $k\in\{-\ol{m},-\ol{m}+1,\cdots,0\}$.

In either case 1 or case 3, $$\mathcal{J}_A\left(m_0\big|\left\{a_B^*,a_B^\text{br}\right\}\right)>v_A\left(m_0\big|\left\{a_A^*,a_B^*\right\}\right).$$
In other words, firm $A$ can achieve a strictly higher performance measure by deviating from $a_A^*$ to $a_B^*$. However, that it chooses not to contradicts the rational agent assumption. Therefore the premise that $\exists m_0\in\mathcal{M}$ such that $v_A(m_0)\leq v_B(m_0)$ is false. 
\end{proof}
%%%%%%%%%%%%%%%%%%%%%%%%%%%%%%%%%%
\subsection{Proof of Lemma \ref{lem:rdi}}
\begin{proof}
Firstly, if $a_A^*(-1) \leq a_B^*(-1)$, it must be that $a_A^*(0) < a_B^*(0)$. To see this, notice that from Corollary \ref{cor:mpe},
\begin{align}
    &v_f(1) - v_f(0) \\
    =&\dfrac{1}{\lambda a^*_{-f}(-1)+\rho+h}
    \left[\pi_f(1) + \left(\lambda a_{-f}^*(-1)+h\right)v_f(0) 
    - \left(\lambda a_{-f}^*(-1)+h\right)v_f(0) -\rho v_f(0)\right] \nonumber \\
    =& \dfrac{1}{\lambda a_{-f}^*(-1)+\rho+h}\left[\pi_f(1)-\rho v_f(0)\right]. \label{app:v1-v0}
\end{align}
Suppose $a_A^*(-1) \leq a_B^*(-1)$, then from Corollary~\ref{cor:mpe}, Lemma \ref{lem:valFcn}, Equation~(\ref{app:v1-v0}) and Assumption~\ref{asm:pi} that $\pi_A(m)=\pi_B(m)$ $\forall m\in\mathcal{M}$,
\begin{align}
    \dfrac{\partial\psi_A\left(a_A^*(0)\right)}{\partial a_A}
    =&\lambda[v_A(1)-v_A(0)] \nonumber \\
    =& \dfrac{\lambda}{\lambda a_B^*(-1)+\rho+h}\left[\pi_A(1)-\rho v_A(0)\right] \nonumber \\
    <&
    \dfrac{\lambda}{\lambda a_A^*(-1)+\rho+h}\left[\pi_B(1)-\rho v_B(0)\right] \nonumber \\
    =& \lambda[v_B(1)-v_B(0)] \nonumber \\
    =&
    \dfrac{\partial\psi_B\left(a_B^*(0)\right)}{\partial a_B}. \label{app:neqFocRde}
\end{align}
Since $\dfrac{\partial\psi_A(a)}{\partial a_A}<\dfrac{\partial\psi_B(a)}{\partial a_B}$, inequality (\ref{app:neqFocRde}) implies $a_A^*(0)<a_B^*(0)$. Now that $a_A^*(-1) \leq a_B^*(-1)$ and $a_A^*(0) < a_B^*(0)$, for either $m=-1$ or $m=0$, 
$$\dfrac{d\psi_A\left(a_A^*(m)\right)}{da_A}<\dfrac{d\psi_B\left(a_B^*(m)\right)}{da_B}.$$ 
Therefore
\begin{align}
    &[v_A(1)-v_B(1)] - [v_A(-1)-v_B(-1)] \nonumber \\
    =& [v_A(1)-v_A(-1)] - [v_B(1)-v_B(-1)] \nonumber \\
    =& [v_A(1) - v_A(0) + v_A(0) - v_A(-1)] 
    - [v_B(1) - v_B(0) + v_B(0) - v_B(-1)] \nonumber \\
    =& \left(\dfrac{\partial\psi_A\left(a_A^*(0)\right)}{\partial a_A} + \dfrac{\partial\psi_A\left(a_A^*(-1)\right)}{\partial a_A}\right)
    -
    \left(\dfrac{\partial\psi_B\left(a_B^*(0)\right)}{\partial a_B} + \dfrac{\partial\psi_B\left(a_B^*(-1)\right)}{\partial a_B}\right) < 0. \label{app:valFcnDid}
\end{align}
On the other hand, from Corollary \ref{cor:mpe}, when $\ol{m}=1$, 
\begin{align}
    v_A(1) - v_B(1) =& \dfrac{1}{\lambda a_B^*(-1) + \rho + h}
    \left[\pi(1) + \left(\lambda a_B^*(-1) + h\right)v_A(0)\right] \nonumber \\
    &- \dfrac{1}{\lambda a_A^*(-1) + \rho + h}
    \left[\pi(1) + \left(\lambda a_A^*(-1) + h\right)v_B(0)\right]; \label{app:valFcnDiffA} \\
    v_A(-1) - v_B(-1) =& \dfrac{1}{\lambda a_A^*(-1) + \rho + h}
    \left[\pi(-1) - \psi_A\left(a_A^*(-1)\right) + \left(\lambda a_A^*(-1) + h\right)v_A(0)\right] \nonumber \\ 
    &- \dfrac{1}{\lambda a_B^*(-1) + \rho + h}
    \left[\pi(-1) - \psi_B\left(a_B^*(-1)\right) + \left(\lambda a_B^*(-1) + h\right)v_B(0)\right]. \label{app:valFcnDiffB}
\end{align}
Equations~(\ref{app:valFcnDiffA}) and (\ref{app:valFcnDiffB}) imply
\begin{align}
    &[v_A(1)-v_B(1)] - [v_A(-1)-v_B(-1)] \nonumber \\
    =& \dfrac{1}{\lambda a_B^*(-1) + \rho + h}
    \left[\pi(1) + \pi(-1) + \psi_B\left(a_B^*(-1)\right) + 
    \left(\lambda a_B^*(-1) + h\right)\left(v_A(0) + v_B(0)\right)\right] \nonumber \\
    &- \dfrac{1}{\lambda a_A^*(-1) + \rho + h}
    \left[\pi(1) + \pi(-1) + \psi_A\left(a_A^*(-1)\right) + 
    \left(\lambda a_A^*(-1) + h\right)\left(v_A(0) + v_B(0)\right)\right]
    \label{app:valFcnDidAlt}
\end{align}
From $\psi_f(0)=0$ in Assumption \ref{asm:psi} and $\dfrac{\partial\psi_A(a)}{\partial a_A}<\dfrac{\partial\psi_B(a)}{\partial a_B}$ for all $a\geq 0$,  $$\forall a>0, \psi_A(a)<\psi_B(a).$$ 
Therefore, in Equation~(\ref{app:valFcnDidAlt}), when arrival rate multiplier $\lambda>0$ is small enough, 
$$[v_A(1)-v_B(1)] - [v_A(-1)-v_B(-1)]>0.$$
This contradicts inequality (\ref{app:valFcnDid}). Hence, $a_A^*(-1) \leq a_B^*(-1)$ is negated.

Now that $a_A^*(-1)>a_B^*(-1)$, similar to inequality (\ref{app:neqFocRde}), when the discount factor $\rho>0$ is small enough, 
$$\dfrac{\partial\psi_A\left(a_A^*(0)\right)}{\partial a_A}>\dfrac{\partial\psi_B\left(a_B^*(0)\right)}{\partial a_B},$$
which implies $a_A^*(0)>a_B^*(0)$.
\end{proof}
%%%%%%%%%%%%%%%%%%%%%%%%%%%%%%%%%%%%
\subsection{Proof of Lemma \ref{lem:compValFcn}}
\begin{proof}
Define the lower contour set $\mathcal{C}_f\left(v\mid d_f,d_{-f}\right)$ as the set of all strategy profiles $\left\{a_A(t),a_B(t)\right\}_{t=0}^\infty$ by which the performance measure of firm $f$ is no greater than $v$:
\begin{align}
    \mathcal{C}_f\left(v\mid d_f,d_{-f}\right)\coloneqq \left\{\left\{a_A(t),a_B(t)\right\}_{t=0}^\infty:\mathcal{J}_f\left(d_f,d_{-f}\mid\left\{a_f(t),a_{-f}(t)\right\}_{t=0}^\infty\right)\right\}
\end{align}
It is important that both contour sets $\mathcal{C}_A$ and $\mathcal{C}_B$, when non-empty, have their elements in the same order of the strategies of firms $A$ and $B$, otherwise any operation of these two sets is meaningless.

Fix $d_A$ and $d_B$, suppose firm $f$'s value function $v_f$ is defined at $(d_f,d_{-f})$ and $(d_f+1,d_{-f}+1)$. By Assumption \ref{asm:piExt}, for any arbitrary strategy profile $\left\{\widetilde{a}_A(t),\widetilde{a}_B(t)\right\}_{t=0}^\infty$, $$\mathcal{J}_f\left(d_f,d_{-f}\mid\left\{\widetilde{a}_A(t),\widetilde{a}_B(t)\right\}_{t=0}^\infty\right)>\mathcal{J}_f\left(d_f+1,d_{-f}+1\mid\left\{\widetilde{a}_A(t),\widetilde{a}_B(t)\right\}_{t=0}^\infty\right).$$ 
Therefore, 
\begin{equation}\label{app:contourSub}
    \mathcal{C}_f\left(v_f(d_f+1,d_{-f}+1)\mid d_f,d_{-f}\right)\subsetneq \mathcal{C}_f\left(v_f(d_f+1,d_{-f}+1)\mid d_f+1,d_{-f}+1\right)
\end{equation}
and
\begin{equation}\label{app:contourCap}
    \partial\mathcal{C}_f\left(v_f(d_f+1,d_{-f}+1)\mid d_f,d_{-f}\right)\cap \partial\mathcal{C}_f\left(v_f(d_f+1,d_{-f}+1)\mid d_f+1,d_{-f}+1\right) = \emptyset,
\end{equation}
where $\partial \mathcal{C}$ denotes the boundary of set $\mathcal{C}$. By the definition of equilibrium strategy profiles,
\begin{align*}
    &\left\{a_A^*(t),a_B^*(t)\mid d_A+1,d_B+1\right\}_{t=0}^\infty \in \mathcal{C}_f\left(v_f(d_f+1,d_{-f}+1)\mid d_f+1,d_{-f}+1\right), \\
    &\left\{a_A^*(t),a_B^*(t)\mid d_A+1,d_B+1\right\}_{t=0}^\infty \notin \mathcal{C}_f\left(v_f(d_f+1,d_{-f}+1)\mid d_f,d_{-f}\right),
\end{align*}
which implies that
\begin{equation}\label{app:contourMinus}
    \mathcal{C}_f\left(v_f(d_f+1,d_{-f}+1)\mid d_f+1,d_{-f}+1\right)\setminus \mathcal{C}_f\left(v_f(d_f+1,d_{-f}+1)\mid d_f,d_{-f}\right) \neq \emptyset.
\end{equation}
Now discuss whether 
$$\left\{a_A^*(t),a_B^*(t)\mid d_A,d_B\right\}_{t=0}^\infty\in\mathcal{C}_f\left(v_f(d_f+1,d_{-f}+1)\mid d_f,d_{-f}\right).$$ 
Suppose not, then by the definition of lower contour set, $$v_f(d_f,d_{-f})>v_f(d_f+1,d_{-f}+1),$$ 
and the proof thus finishes.

If it does, then by (\ref{app:contourCap}) and (\ref{app:contourMinus}), firm $f$ can deviate to any strategy in the non-empty difference set in (\ref{app:contourMinus}), where for any strategy $-f$ can choose, firm $f$ will end up with a strictly higher performance measure than $v_f(d_f,d_{-f})$, which contradicts that $\left\{a_A^*(t),a_B^*(t)\mid d_A,d_B\right\}_{t=0}^\infty$ consists of the optimal strategies for both firms.
\end{proof} 
%%%%%%%%%%%%%%%%%%%%%%%%%%%%%%%%%%%%
\subsection{Proof of Proposition \ref{prp:asymProb}}
\begin{proof}
By the definition of $Q$-matrix and Lemma \ref{lem:rdi}, the transition rate matrix $Q$ is
\begin{equation}\label{eqn:QSpec}
    Q = 
    \begin{bmatrix}
        -\left(\lambda a_A^*(-1) + h\right) & \lambda a_A^*(-1) + h & 0 \\
        \lambda a_B^*(0) & -\left(a_A^*(0) + a_B^*(0)\right) & \lambda a_A^*(0) \\
        0 & \lambda a_B^*(-1) + h & -\left(\lambda a_B^*(-1) + h\right)
    \end{bmatrix}.
\end{equation}
The \textit{invariant measure} $\xi$ is the solution to $\xi Q=0$, whose components satisfy
\begin{equation}\label{eqn:invMeasComp}
    \dfrac{\xi(1)}{\xi(-1)}=\dfrac{\lambda a_A^*(0)}{\lambda a_B^*(0)}\cdot \dfrac{\lambda a_A^*(-1) + h}{\lambda a_B^*(-1) + h} > 1.
\end{equation}
By Definitions \ref{def:matP}, \ref{def:statDist} and Proposition \ref{prp:statDist}, 
$$\mu(1)=\xi(1)\left(\lambda a_A^*(-1) + h\right)>\xi(-1)\left(\lambda a_B^*(-1) + h\right)=\mu(-1).$$
\end{proof}
%%%%%%%%%%%%%%%%%%%%%%%%%%%%%%%%%%%%
\subsection{Proof of Proposition \ref{prp:shkLarge}}
\begin{proof}
Suppose the state prior to the shock is $(d_A,d_B)$. For either firm $f\in\{A,B\}$, by Lemma \ref{lem:compValFcn}, both $v_f(d_f+k-1,d_{-f}+k)$ and $v_f(d_f+k,d_{-f}+k)$ are monotonically decreasing in $k$ and bounded below by zero. Therefore, as $\delta\rightarrow 0^+$, 
$\displaystyle \lim_{k\rightarrow \infty}v_f(d_f+k-1,d_{-f}+k)$ and $\displaystyle \lim_{k\rightarrow \infty}v_f(d_f+k,d_{-f}+k)$ exists and are equal. This implies that either
$$\displaystyle\lim_{k\rightarrow \infty}\left[v_f(d_f+k-1,d_{-f}+k)-v_f(d_f+k,d_{-f}+k)\right]=0,$$ 
or $\forall \varepsilon>0$, $\exists \widetilde{D}(\varepsilon)$ such that when $k\geq \widehat{D}(\varepsilon)$,
$$v_f(d_f+k-1,d_{-f}+k)-v_f(d_f+k,d_{-f}+k)<\varepsilon,$$
By Proposition \ref{prp:existExt} and that R\&D cost function $\psi_f$ is strictly increasing (Assumption \ref{asm:psi}), this implies that for either firm, for a large enough $k$, the equilibrium R\&D effort satisfies $$a_f^*(d_f,d_{-f})>a_f^*(d_f+k,d_{-f}+k).$$
\end{proof} 
%%%%%%%%%%%%%%%%%%%%%%%%%%%%%%%%%%%%
\subsection{Proof of Proposition \ref{prp:shkSmall}}
\begin{proof}
Without loss of generality, suppose $\displaystyle\lim_{s\rightarrow t^-}d_A(s)=0$. We first discuss the case in which $\displaystyle\lim_{s\rightarrow t^-}d_B(s)<\ol{m}$. By updating rule (\ref{eqn:transRule}) and first-order condition (\ref{eqn:existExtPol}), we have both
$$\displaystyle\lim_{s\rightarrow t^-}a_A^*(s)=\psi_A^{-1}\left(\lambda\left[v_A(0,d_B+1)-v_A(0,d_B)\right]\right)$$ and 
$$a_A^*(t)=\psi_A^{-1}\left(\lambda\left[v_A(0,d_B+1)-v_A(1,d_B+1)\right]\right).$$ 
By Lemma \ref{lem:compValFcn}, 
$$v_A(0,d_B)>v_A(1,d_B+1).$$ 
By the strict monotonicity of $\psi_A$ in Assumption \ref{asm:psi},  $$a_A^*(t)>\displaystyle\lim_{s\rightarrow t^-}a_A^*(s).$$
Now consider the case $\displaystyle\lim_{s\rightarrow t^-}d_B(s)=\ol{m}$. By boundary condition (\ref{eqn:existExtPolBnd}), $$\displaystyle\lim_{s\rightarrow t^-}a_A^*(s)=0.$$ 
By first-order condition (\ref{eqn:existExtPol}), 
$$a_A^*(t)>0.$$ 
Therefore 
$$a_A^*(t)>\displaystyle\lim_{s\rightarrow t^-}a_A^*(s).$$ 
\end{proof}
%%%%%%%%%%%%%%%%%%%%%%%%%%%%%%%%%%%%
%%%%%%%%%%%%%%%%%%%%%%%%%%%%%%%%%%%%
\newpage
\section{Computation Appendix} \label{Computation Appendix}
\setcounter{equation}{0}
\numberwithin{equation}{section}
\renewcommand{\theequation}{C.\arabic{equation}}
%%%%%%%%%%%%%%%%%%%%%%%%%%%%%%%%%%%%

%%%%%%%%%%%%%%%%%%%%%%%%%%%%%%%%%%%%
\subsection{Value Function Iteration} \label{Computation Appendix: VFI}
This subsection describes the procedures to solve the baseline model numerically. 
\begin{enumerate}
    \item Set initial guess $\boldsymbol{v}_0=\{v_{0,A}(m),v_{0,B}(m)\}_{m\in\mathcal{M}}$ for all $m\in\mathcal{M}$ and $f\in\{A,B\}$. Our initial guess is $v_f(m)=0$ .
    \item Compute policy function $\boldsymbol{a}_0^*=\{a_{0,A}^*(m),a_{0,B}^*(m)\}_{m\in\mathcal{M}}$ according to (\ref{eqn:aFoc}) and (\ref{eqn:aMax}).
    \item Update value function $v_{1,f}(m)$ using $\boldsymbol{v}_0$ and $\boldsymbol{a}_0^*$ on the right-hand side in (\ref{eqn:valFcnBsl}) and (\ref{eqn:vMax}) to get $\boldsymbol{v}_1$.
    \item Keep iterating until the sequence of value functions $\{\boldsymbol{v}_k\}_{k=0}^\infty$ converges.
\end{enumerate}
%%%%%%%%%%%%%%%%%%%%%%%%%%%%%%%%%%%%
\subsection{GMM Estimation} \label{Computation Appendix: GMM}
Denote $\theta:=(\alpha,\gamma,h)$. The numerical solution of the policy functions 
$$\left\{a_A^*(\Delta n_A\mid\theta),a_B^*(\Delta n_B\mid\theta)\right\}_{\Delta n_A=-\Delta n_B\in\mathcal{M}},$$
is conditional on $\theta$. By Proposition \ref{prp:statDist}, the stationary distribution $\mu(m \mid \theta)$ is also a function on parameter set $\theta$. We construct selected moments using the stationary distribution, and calculate their weighted distance to their counterparts in the data. This distance is then minimized over the parameter space to pin down the value of $\theta$. The following is a step-by-step summary of our approach regarding the GMM estimation. 
\begin{enumerate}
    \item The first moment is the expected ratio of the low-cost firm's and high-cost firm's R\&D efforts:
    \begin{equation}\label{app:mdlMmt1}
        \phi_1 = \ex \left[\dfrac{a_A^*(\Delta n_A)}{a_B^*(\Delta n_B)}\right].
    \end{equation}
    The second one is the expected ratio of firm values between the two firms:
    \begin{equation}\label{app:mdlMmt2}
        \phi_2 = \ex \left[\dfrac{\left(v_A(\Delta n_A)\right)}{\left(v_B(\Delta n_B)\right)}\right].
    \end{equation}
    The last target moment is the expected ratio of the profits:
    \begin{equation}\label{app:mdlMmt3}
        \phi_3 = \ex \left[\dfrac{\pi_A(\Delta n_A)}{\pi_B(\Delta n_B)}\right],
    \end{equation}
    where the profit function $\pi$ is determined by (\ref{eqn:piSpec1}) and (\ref{eqn:piSpec2}).
    \item 
    Conditional on $\theta$, solve the baseline model numerically as illustrated in Appendix~\ref{Computation Appendix: VFI}, and compute the stationary distribution $\mu(\theta)$ based on the transition rate matrix $Q$ and jump matrix $P$ in Equation~(\ref{eqn:matP}). Then we compute the moments using $\mu(m|\theta)$ as follows:
    \begin{align}
        \phi_1(\theta) =& \sum_{m=-\ol{m}}^{\ol{m}}\mu(m|\theta)\dfrac{a_A^*(m|\theta)}{a_B^*(-m|\theta)}; \label{app:staMmt1} \\
        \phi_2(\theta) =& \sum_{m=-\ol{m}}^{\ol{m}}\mu(m|\theta)\dfrac{v_A(m|\theta)}{v_B(-m|\theta)}; \label{app:staMmt2} \\
        \phi_3(\theta) =& \sum_{m=-\ol{m}}^{\ol{m}}\mu(m|\theta)\dfrac{\pi_A(m|\theta)}{\pi_B(-m|\theta)}. \label{app:staMmt3}        
    \end{align}
    \item Compute the corresponding moments from data, denoted as  $\Phi=(\Phi_1,\Phi_2,\Phi_3)$. We use R\&D expenditure (RDE) to proxy for the R\&D cost $\psi$, market value, $mkv$, for firm value and gross profit $gp$ for the profit.
    \begin{align}
        \Phi_1 =& \dfrac{1}{JT}\sum_{j,t}
        \dfrac{\sum_{i}rde_{i,j,t}\times \mathds{1}\{c_{i,j,t}^8=1\}}{\sum_{i}rde_{i,j,t}\times \mathds{1}\{c_{i,j,t}^3=1\}}; \label{app:empMmt1} \\
        \Phi_2 =& \dfrac{1}{JT}\sum_{j,t}
        \dfrac{\sum_{i}mkv_{i,j,t}\times \mathds{1}\{c_{i,j,t}^8=1\}}{\sum_{i}mkv_{i,j,t}\times \mathds{1}\{c_{i,j,t}^3=1\}}; \label{app:empMmt2} \\
        \Phi_3 =& \dfrac{1}{JT}\sum_{j,t}
        \dfrac{\sum_{i}gp_{i,j,t}\times \mathds{1}\{c_{i,j,t}^8=1\}}{\sum_{i}gp_{i,j,t}\times \mathds{1}\{c_{i,j,t}^3=1\}}; \label{app:empMmt3}
    \end{align}
    Take the calculation of $\Phi_1$ for example. Firstly for each industry-year cell $(j,t)$, take the sum of RDE over the eighth and third decile groups, respectively. Secondly, compute the ratio of these two sums. This is equivalent to computing the ratio of the means of the RDE from these two decile groups, because the number of observation in each group is equal by definition. Finally, compute the mean of this ratio across all industry-year cells. This is the sample counterpart of moment (\ref{app:mdlMmt1}), the expected ratio of R\&D costs. The similar can be said for (\ref{app:empMmt2}) and (\ref{app:empMmt3}).
    \item Let $\widehat{W}$ be the covariance matrix of variables $\dfrac{\sum_{i}rde_{i,j,t}\times \mathds{1}\{c_{i,j,t}^8=1\}}{\sum_{i}rde_{i,j,t}\times \mathds{1}\{c_{i,j,t}^3=1\}}$, $\dfrac{\sum_{i}mkv_{i,j,t}\times \mathds{1}\{c_{i,j,t}^8=1\}}{\sum_{i}mkv_{i,j,t}\times \mathds{1}\{c_{i,j,t}^3=1\}}$ and $\dfrac{\sum_{i}gp_{i,j,t}\times \mathds{1}\{c_{i,j,t}^8=1\}}{\sum_{i}gp_{i,j,t}\times \mathds{1}\{c_{i,j,t}^3=1\}}$, invert it to get $\widehat{W}^{-1}$. The weighted distance between $\phi(\theta)=\left(\phi_1(\theta),\phi_2(\theta),\phi_3(\theta)\right)$ and $\Phi$ is a function of the parameter set $\theta$:
    \begin{equation}\label{app:dst}
        G(\theta) = \left(\phi(\theta)-\Phi\right)^\prime \widehat{W}^{-1} \left(\phi(\theta)-\Phi\right).
    \end{equation}
    This estimator is minimized over the parameter space $\left\{(\alpha,\gamma,h):0<\alpha<1,\gamma>1,h>0\right\}$, and the minimizer is the estimated values. Since the dimension is low, we use the grid search method with multiple starting points.   
\end{enumerate}
%%%%%%%%%%%%%%%%%%%%%%%%%%%%%%%%%%%%
\subsection{Model Simulations} \label{Computation Appendix: Simulations}
\setcounter{equation}{0}
\numberwithin{equation}{section}
\renewcommand{\theequation}{D.\arabic{equation}}
First discretize the time horizon to $T=1000$ periods with interval $\Delta t = 0.05$ and set the number of simulations $S$ to be 1000000, thus $20$ periods in simulation correspond to one year -- we need a small $\Delta t$ to approximate the Poisson process of innovation. Create $\widehat{d}_{A,S}$ and $\widehat{d}_{B,S}$, two matrices with dimensions $S\times T$, to store the values of distances to the technology frontier for each simulation $s$. Create $\widehat{a}_{A,S}$ and $\widehat{a}_{B,S}$ to store equilibrium R\&D efforts in each simulation $s$.

\subsubsection*{Updating State Variables and Tech Gap}

In each simulation $s$, firm $f$'s optimal R\&D effort at time $t$ is $a_f^*\left(\widehat{d}_{f,s}(t),\widehat{d}_{-f,s}(t)\right)$. By Markov assumption, starting from $t=2$, the probability distribution of the state $\left(\widehat{d}_{A,s}(t),\widehat{d}_{B,s}(t)\right)$ is determined by the last period's state $\left(\widehat{d}_{A,s}(t-1),\widehat{d}_{B,s}(t-1)\right)$ and optimal R\&D efforts. 

Recall that firm $f$'s location on the technology ladder, $N_f(t)$, is determined by its R\&D success, which is governed by a Poisson Process with arrival rate 
\begin{equation}\label{app:arrRate}
    \lambda_f(t) = \lambda a_f^*(t) + h\times \mathds{1}\{N_f(t)<N_{-f}(t)\},
\end{equation}
where $a_f^*(t)$ denotes $f$'s equilibrium R\&D effort. Without negative profit shocks, $a_f^*(t)=a_f^*\left(\Delta N_f(t)\right)$ by Corollary \ref{cor:mpe}. The transitions of the Poisson process $N_f(t)$ in time interval $[t,t+\Delta t]$ are approximately the following:
\begin{align}
    \mathds{P}\left(N_f(t+\Delta t)-N_f(t)=0\right) =& \exp[-\lambda_f(t)\Delta t]\approx 1-\lambda_f(t)\Delta t, \label{app:approx1} \\
    \mathds{P}\left(N_f(t+\Delta t)-N_f(t)=1\right) =& \lambda_f(t)\Delta t\exp[-\lambda_f(t)\Delta t]\approx \lambda_f(t)\Delta t, \label{app:approx2} \\    
    \mathds{P}\left(N_f(t+\Delta t)-N_f(t)\geq 2\right) =& 1-(1+\lambda_f(t)\Delta t)\exp[-\lambda_f(t)\Delta t]\approx 0. \label{app:approx3}
\end{align}

To program Equation~(\ref{app:approx1}) -- (\ref{app:approx3}), denote $\widehat{N}$ as the discrete analogue of the continuous-time process $N$ as in Section \ref{sec:sim}, . The updating of $\widehat{d}_f(t)$ follows the rules specified by Equation~(\ref{eqn:transRule}). Generate a length $T$ scalar $p_f(t)$ to store the jumping probability $\lambda_f(t)\Delta t$. For each period $t$, draw $U_f(t)$ from i.i.d. uniform distribution $U[0,1]$. For $t\geq2$, 
\[
\widehat{N}_f(t)=
\begin{cases}
\widehat{N}_f(t-1)+1, & U_f(t)<p_f(t-1); \\
\widehat{N}_f(t-1), & \text{otherwise}.
\end{cases}
\]
As such, a success arrives with probability $p_f(t-1)$ at time $t$.

The transition of technology gap $\lvert \widehat{N}_A(t)-\widehat{N}_B(t)\rvert$ is inferred from the transitions of firms' respective transitions. Regarding the automatic catching up assumption, we add to the updating rule that if $\Delta \widehat{N}_f(t-1)=\ol{m}$, $U_f(t)<p_f(t-1)$ and $U_{-f}(t)\geq p_{-f}(t-1)$, then $\widehat{N}_f(t)=\widehat{N}_f(t-1)+1$ and $\widehat{N}_{-f}(t)=\widehat{N}_{-f}(t-1)+1$. In other word,when the leader is at the maximum gap and innovates, the laggards automatically advances on step as well along the technology ladder, so that the technology gap remains unchanged, even though no innovation is achieved simultaneously by the latter.

\subsubsection*{Simulating the Economy}

For each repetition $s$, start with $\widehat{N}_{A,s}(1)=\widehat{N}_{B,s}(1)=1$. Accordingly, firms are neck--and--neck on the frontier in the first period, i.e., $\widehat{d}_{A, s}(1)=\widehat{d}_{B, s}(1)=0$.

For $t=1,\cdots,899$, the technology gap is $m_f = d_{-f} - d_f$ and distance to frontier is 0. A negative profit shock with degree $D=4$ occurs at $t=900$. Both firms are pushed backward from the technology frontier by four steps simultaneously, while their relative positions remain unchanged. In the code, after the realization of the state $\left(\widehat{d}_{A,s}(900),\widehat{d}_{B,s}(900)\right)$, we change it to $\left(\widehat{d}_{A,s}(900)+4,\widehat{d}_{B,s}(900)+4\right)$. 

The shock is set to hit the economy at $t=900$ so that there is sufficient time for the stochastic processes to evolve and reach the limiting distribution. We set $D=4$ so that conditional on $\delta=0.05$ in Equation~(\ref{eqn:piExt}), the negative profit shock lowers about $20\%$ of the gross profit at the industry level. 

\subsubsection*{Computing the Responses}

Repeat the procedure $S$ times. The impulse responses of expected R\&D efforts to the shock is approximated by the mean of the simulation paths as follows:
\begin{equation}\label{eqn:simPol}
    \widehat{a}_f(t) = \dfrac{1}{S}\sum_{ b=1}^{S}\widehat{a}_{f,s}(t),
\end{equation}
For $t\geq 900$, it shows the simulated IRF of firm $f$'s R\&D effort to the shock. Similarly, the simulated technology gap of firm $f$, $\Delta \widehat{n}_f$, is calculated as
\begin{equation}\label{eqn:simGap}
    \Delta \widehat{n}_f(t) = \dfrac{1}{S}\sum_{b=1}^{S}\Big(\widehat{d}_{-f,s}(t)-\widehat{d}_{f,s}(t)\Big).
\end{equation}

%The algorithm below summarizes the procedures. 
%\begin{algorithm}[h]
% \SetAlgoLined
 %\KwData{this text}
 %\KwResult{how to write algorithm with \LaTeX2e }
% Initialization: $\widehat{d}_{A, S}, \widehat{d}_{B, S}, \widehat{a}_{A, S}, \widehat{a}_{B, S}$\;
% \For{$s = 1, \dots, S$}{
%  read current\;
% \eIf{understand}{
%  go to next section\;
% current section becomes this one\;
% }{
% go back to the beginning of current section\;
% }
%}
%\caption{Simulation Procedures}
%\end{algorithm}

\end{appendices}
\end{document}